%% file: main.tex
\documentclass[a4paper,11pt]{article}
\pdfoutput=1 

\usepackage{jinstpub} 
\usepackage{verbatim}

\usepackage{amsmath}

\usepackage{inputenc,graphicx}
\usepackage{siunitx}
\usepackage{multirow}
\usepackage{makecell}

\newcommand{\ipreamp}{I_{\it preamp}} 
 
\newcommand{\sigdtoa}{\sigma_{\Delta_{\it TOA}}} 
\newcommand{\dtoa}{\Delta_{\it TOA}} 
\newcommand{\sigdut}{\sigma_{\it DUT}}
\newcommand{\siglgzero}{\sigma_{\it LGAD0}}
\newcommand{\siglgone}{\sigma_{\it LGAD1}}

\title{Testbeam Results of the Picosecond Avalanche Detector Proof-Of-Concept Prototype}

\author[a,1]{G. Iacobucci,\note{Corresponding author.}}
\author[a]{S. Zambito,}
\author[a]{M. Milanesio,}
\author[a]{T. Moretti,}
\author[a]{J. Saidi,}
\author[a,b]{L. Paolozzi,} 
\author[a]{M. Munker,}
\author[a]{R. Cardella,}
\author[a]{F. Martinelli,}
\author[a,b]{A. Picardi,}
\author[c]{H. Rücker,}
\author[c]{A. Trusch,}
\author[a]{P. Valerio,}
\author[a]{F. Cadoux,}
\author[a,2]{R. Cardarelli\note{Also at INFN Section of Roma Tor Vergata, Via della ricerca scientifica 1, Roma, Italy.},}
\author[a]{S. Débieux,}
\author[a]{Y. Favre,}
\author[a]{C. A. Fenoglio,}
\author[a]{D. Ferrere,}
\author[a]{S. Gonzalez-Sevilla,}
\author[a]{Y. Gurimskaya,}
\author[a,b]{R. Kotitsa,}
\author[a]{C. Magliocca,}
\author[a,b]{M. Nessi,}
\author[a]{A. Pizarro-Medina,}
\author[a]{J. Sabater Iglesias,}
\author[a]{and M. Vicente Barreto Pinto.}
\affiliation[a]{D\'epartement de Physique Nucl\'eaire et Corpusculaire (DPNC),
University of Geneva, 24 Quai Ernest-Ansermet, CH-1211 Geneva 4, Switzerland}

\affiliation[b]{CERN, CH-1211 Geneva 23, Switzerland}

\affiliation[c]{IHP — Leibniz-Institut für innovative Mikroelektronik, Im Technologiepark 25, Frankfurt (Oder), Germany}

\emailAdd{giuseppe.iacobucci@unige.ch}

\abstract{The proof-of-concept prototype of the Picosecond Avalanche Detector, a multi-PN junction monolithic silicon detector with continuous gain layer deep in the sensor depleted region,  was tested with a beam of 180 GeV pions at the CERN SPS.
The prototype features   low noise and fast SiGe BiCMOS frontend electronics and hexagonal pixels with 100 µm pitch.
At a sensor bias voltage of 125 V, the detector provides full efficiency and average time resolution of 30, 25 and 17 ps in the overall pixel area 
for a power consumption of 0.4, 0.9 and 2.7 W/cm$^2$, respectively.
In this first prototype the time resolution depends significantly on the distance from the  center of the pixel, varying  at the highest power consumption measured between 13 ps at the  center of the pixel and 25 ps in the inter-pixel region.
}


\begin{document}

\maketitle
\input{Introduction}
\input{ASIC}

\input{ExperimentalSetup}
\input{Efficiency}

\input{Resolution}

\input{Conclusions}

\acknowledgments
The prototyping of the PicoAD was funded by the  EU H2020 ATTRACT MONPICOAD project under grant agreement 222777, with support by an INNOGAP grant from the University of Geneva. The test and characterization of the prototypes was done in the context of the H2020 ERC Advanced Grant MONOLITH, grant ID: 884447, as well as of the Swiss National Science
Foundation grant number 200020\_188489. 
The authors wish to thank Nicolo Cartiglia and Roberta Arcidiacono for providing the LGAD sensors, Edoardo Charbon and Emanuele Ripiccini for providing the SPAD sensors, Coralie Husi, Javier Mesa, Gabriel Pelleriti and all the technical staff of the University of Geneva and IHP microelectronics.
The authors acknowledge the support of EUROPRACTICE in providing design tools and MPW fabrication services.
\newpage
\bibliographystyle{unsrt}
\bibliography{bibliography.bib}
\end{document}

%% file: Introduction.tex
\section{Introduction}
\label{sec:intro}

Thin silicon sensors are used with great success in high-energy physics experiments for accurate 3D reconstruction of the trajectory of charged particles and the identification of their production vertex. 
In the quest for the next generation of pixel sensors, the particle-physics silicon-detector community is  developing Monolithic Active Pixel Sensors (MAPS)~\cite{peric}, which contain the sensor in the same CMOS substrate utilised for the electronics. MAPS are particularly appealing since they offer all the advantages of an industrial standard processing, avoiding the production complexity and high cost of the bump-bonded hybrid pixel sensors that are commonly used in high-energy physics. So far monolithic pixel sensors with nanosecond time resolution are adopted in the DEPFET~\cite{depfet} implementation in Belle-II~\cite{belle}, while MAPS are used in STAR~\cite{star}, Mu3e~\cite{mu3e} and in the upgraded ALICE tracker~\cite{alice} installed to take data during the LHC Run3. 

With the advent of the  High-Luminosity Large Hadron Collider (HL-LHC) program at CERN,
very precise time measurement  becomes mandatory to cope with the large number of collisions per bunch crossing and be able to extract signals of new physics. The so called 4D event reconstruction~\cite{4D,Sadrozinski_2017}  will be achieved at the HL-LHC by separate detectors~\cite{atlasTDR,cmsTDR}, specialised either in position or in time measurement. This expensive and complex solution to the problem presented by the 200 pileup events per bunch crossing expected at the HL-LHC, will not be sustainable in future hadron-collider programs. 
Monolithic silicon sensors with very high time resolution will enable making 4D measurements better and in a single and cost-effective silicon tracker, and  will also influence how future particle-physics experiments will be designed and constructed.

The production of thin sensors capable to provide excellent position and time resolution at the same time requires to go beyond 
the diode structure of present silicon pixel sensors, that strongly penalises the enormous potential of silicon-based time measurement. 
Indeed, the approximately 30 ps intrinsic limit of diodes was already reached in LGAD sensors with internal gain with pad sizes of 1 mm$^2$~\cite{lgad} and in monolithic sensors without internal gain with 100 µm pixel pitch~\cite{Iacobucci:2021ukp}. 
A promising research is represented by the resistive AC-coupled LGAD approach~\cite{accoupled, accoupled2} in which the $n^+$ implant is resistive and the LGAD is read out by  AC coupling.
An alternative approach is adopted by the TIMESPOT project~\cite{timespot}, that exploits the  3D-sensor structure to produce hybrid timing sensors, intrinsically radiation tolerant. In the version with trenches, using  SiGe BiCMOS discrete electronics, these sensors achieved time resolutions with two Gaussian components: a core with 11 ps and a second component 2.5 times larger. 
The efficiency varies between 80\% for tracks perpendicular to the sensor and full efficiency for track angles larger than 10$^\circ$.

Since several years, this research group is  exploiting the very high speed of SiGe BiCMOS elctronics. 
After a demonstrator with discrete components~\cite{Benoit_2016}, time resolutions of  50 ps was achieved~\cite{hexa_50ps}\cite{Paolozzi_2020} with the first monolithic silicon pixel detector prototype in the SG13G2 IHP 130 nm process \cite{SG13G2} with 100 $\mu$m pixel pitch, without recurring to an avalanche gain mechanism. These first results were reached at a cost in detection efficiency, as a rather high discrimination threshold was imposed to maintain the signal-to-noise ratio demanded by such a remarkable timing performance. Moreover, the procedure used for correcting for signal time walk    limited the timing capabilities of the detector. A second monolithic silicon pixel ASIC prototype was produced in the same 130 nm IHP process, with the objective of surpassing the performance of the previous prototype. A time resolution of \SI{36}{\pico\second} and a detection efficiency of 99.9\% were measured when operating the ASIC at a preamplifier current of \SI{150}{\micro\ampere}~\cite{Iacobucci:2021ukp}.
This result being close to the intrinsic limit of  diodes,  new sensor concepts must be conceived to achieve time resolutions below 10 ps. 

%
In the context of this research, 
the MONOLITH H2020 ERC Advanced project~\cite{monolith} introduces a novel silicon-sensor structure, the Picosecond Avalanche Detector (PicoAD)~\cite{PicoADpatent}, devised to overcome the intrinsic limits of present PN-junction sensors and  provide simultaneous picosecond timing and high spatial resolution in a monolithic implementation. This goal is achieved in the PicoAD by the introduction of a fully depleted multi-junction containing a continuous deep gain layer, which  separates a few µm thick absorption region in which the primary electrons are generated from the region where the electrons drift, and thus decouples the pixelated structure from the continuous gain layer~\cite{picoad_gain}. 
The remarkable  performance expected from this novel sensor, combined with the simplified assembly process and reduced production cost offered by the monolithic implementation in standard CMOS processes, might provide the required breakthrough
and offer a sustainable solution to fundamental researches needing tracking and picosecond time resolution,
among them
future high-energy physics experiments at colliders,
rare-decay experiments featuring active targets or precise tracking,
space-borne astroparticle physics experiments 
and nuclear-physics measurement of lifetimes of excited states in short-lived nuclei.


A PicoAD proof-of-concept prototype was recently produced~\cite{picoad_gain}\cite{monpicoad} using the same foundry masks of the monolithic detector without internal gain studied in~\cite{Iacobucci:2021ukp}.
This first prototype demonstrated that the PicoAD concept works, and the multi-PN junction can be fully depleted and operated in avalanche-gain regime. A detailed description of its working principle, together with the first results on the  characterization at the University of Geneva with  probe station and radioactive sources, is given in~\cite{picoad_gain}.
In the following, an in-depth study of the PicoAD proof-of-concept prototype performance with minimum ionizing particles from a CERN SPS pion beam  is reported.

%% file: ASIC.tex
\section{Description of the ASIC}
\label{sec:asic}

The PicoAD proof-of-concept  prototype studied in this paper is a monolithic silicon detector with 100 µm pixel pitch, produced by IHP  in the SG13G2 130nm SiGe BiCMOS technology. Its novel design characteristic is the separation of the region where the primary charge is produced and amplified from the region dedicated to the signal induction on the readout electrodes~\cite{picoad_gain}. Under the bias voltage, the sensor depletion region spans a multi-junction (NP)$_{\it pixel}$(NP)$_{\it gain}$ structure. The two external junctions operate in inverse polarization: the junction close to the top surface isolates the pixels, while the one close to the bottom edge of the depletion region provides an electric field large enough to generate avalanche gain. The central, direct junction is isolated by the other two junctions, obtaining full depletion. 
\begin{figure}[!htb]
\centering
\includegraphics[width=.99\textwidth,trim=0 0 0 0]{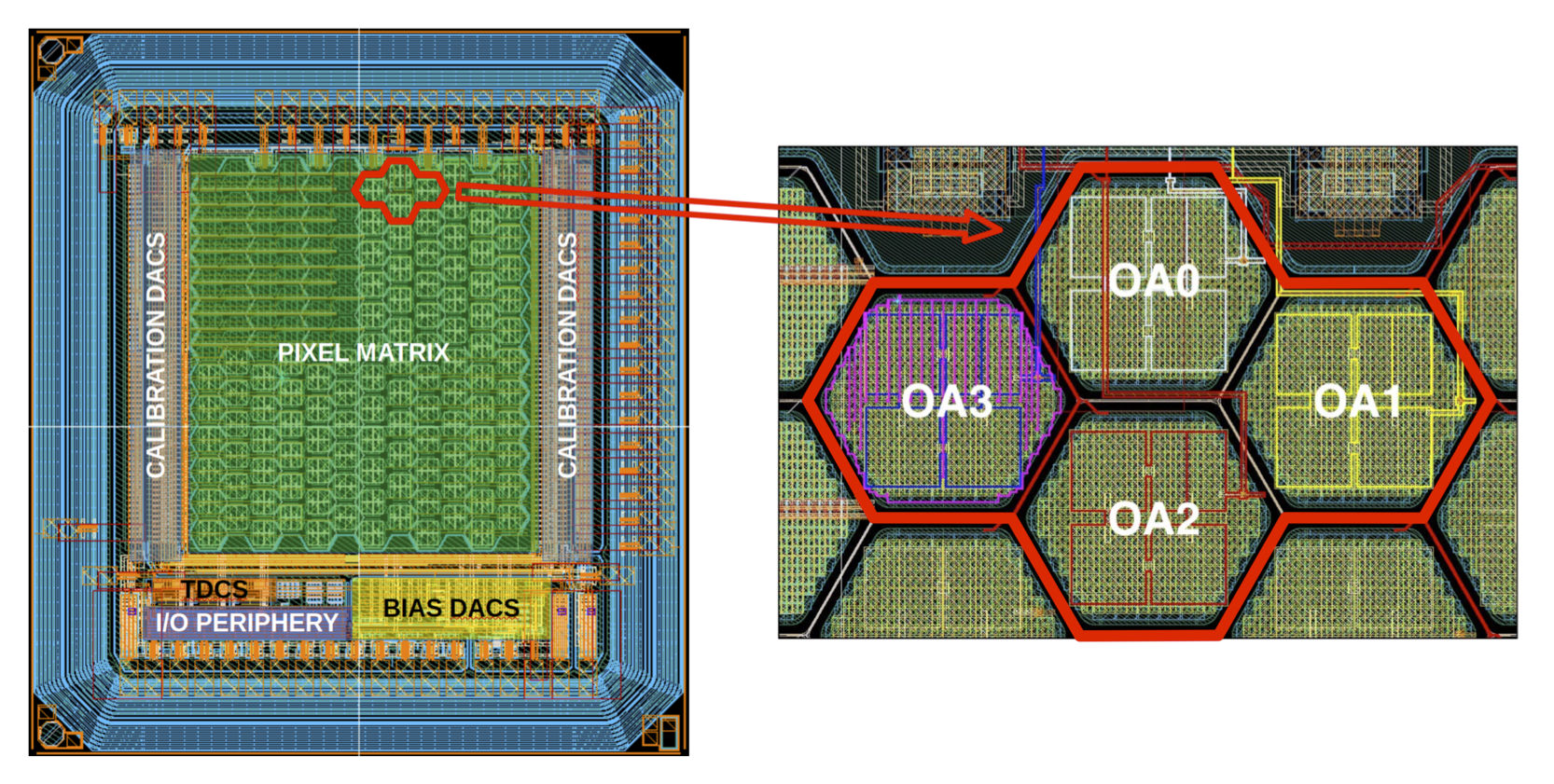}
\caption{\label{fig:chip_layout} (Left) Floorplan of the monolithic pixel detector prototype tested. The ASIC has a total size of 2.3$\times$2.5 mm$^2$. It is divided into four sub-matrices  with  different amplifier design, each containing 6$\times$6 hexagonal pixels of $\SI{65}{\um}$ side. (Right) Magnified detail of the ASIC showing the four analog  pixels  read with two oscilloscopes during the test beam. }
\end{figure}

The prototype was manufactured on boron-doped substrates with a resistivity of 0.1 $\Omega$cm, with a $\SI{5}{\um}$ thick epitaxial (absorption) layer under the gain layer and a $\SI{10}{\um}$ thick epitaxial (drift) layer above the gain layer. For the CMOS processing, the monolithic silicon pixel layout designed for the prototype described in \cite{Iacobucci:2021ukp} was used. The floorplan is shown in Figure~\ref{fig:chip_layout} left. The sensor features a matrix of hexagonal pixels of $\SI{65}{\um}$ side (corresponding to a pitch of approximately $\SI{100}{\um}$), with an inter-pixel distance of $\SI{10}{\um}$. Four of the pixels are connected to charge amplifiers with analog drivers, which are used to measure precisely the charge produced in the sensor. They are hereafter referred to as pixels OA0, OA1, OA2 and OA3; their relative position is shown in Figure~\ref{fig:chip_layout} right.  

%% file: ExperimentalSetup.tex
\section{Experimental Setup and Analysis Criteria}
\label{sec:setup}
The data samples analyzed for measuring the detection efficiency and time resolution of the ASIC prototype were collected at the CERN SPS testbeam facility using a pion beam of \SI{180}{\giga\electronvolt}/c momentum. 

\begin{figure}[!htb]
\centering
\includegraphics[width=.99\textwidth,trim=50 80 50 80,clip]{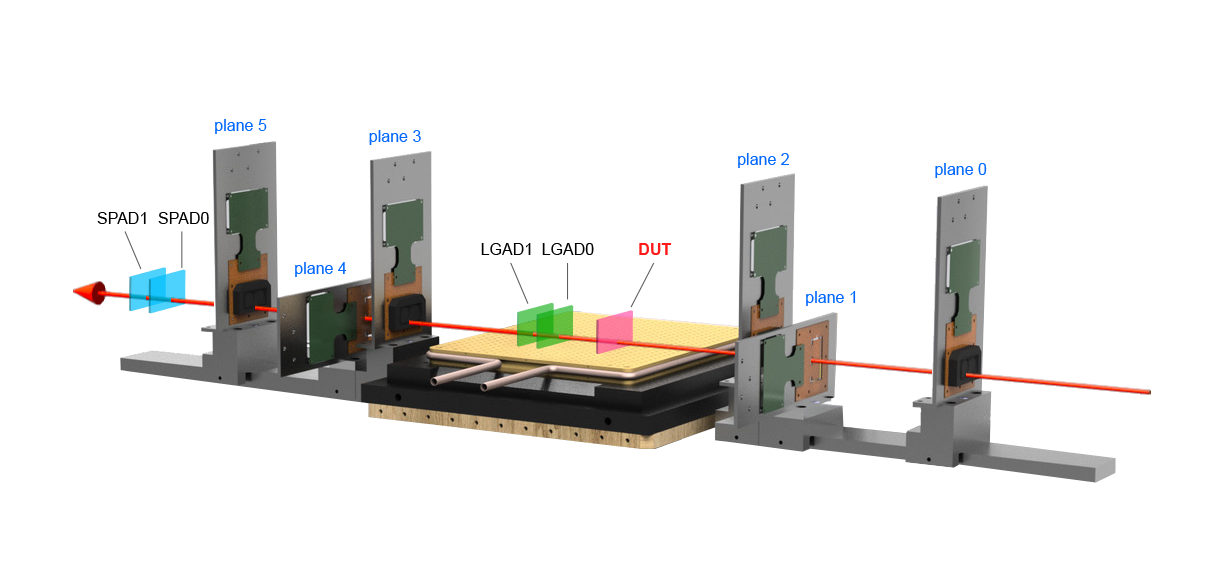}
\caption{\label{fig:Setup} Schematic view of the experimental setup, showing the six planes of the UNIGE FEI4 telescope \cite{FEI4_telescope}, the DUT, the two LGADs and the two SPADs. 
}
\end{figure}

The detector under test (DUT) was interfaced to an aluminum cooling box and operated at a temperature of  $\SI{-20}{\celsius}$. It was positioned, together with two LGAD detectors~\cite{lgad} (named here LGAD0 and LGAD1), in the middle of the six detection planes of the UniGe FEI4 telescope for charged-particle tracking \cite{FEI4_telescope}, as shown in the layout  in Figure~\ref{fig:Setup}. Two SPAD detectors~\cite{charbon} (named here SPAD0 and SPAD1) were installed downstream the telescope to have additional time references of very high precision.

The DUT, the two LGADs and the two SPADs were read by two oscilloscopes, while the telescope planes by a dedicated data acquisition system. Pixel OA0 of the DUT,  SPAD0 and the two LGADs were connected to a first oscilloscope with analog bandwidth of 3 GHz and a sampling rate of 40 GS/s. Pixels OA1, OA2 and OA3 of the DUT and SPAD1 were instead read by a second oscilloscope, with 4 GHz analog bandwidth and a sampling rate of 20 GS/s. The FEI4 telescope provided the trigger to the oscilloscopes. A region of interest of $250 \times 600 $~\si{\um\squared} centered at the OA0 pixel of the DUT was imposed to the first telescope plane, and put in coincidence with the last plane to generate the trigger.

\subsection{Telescope-Track Selection}\label{selection}

Tracks reconstructed with the FEI4 telescope were extrapolated to the DUT plane to determine the position where each pion crossed the sensor. A set of track-quality criteria was designed for selecting a sub-sample of events that maximizes the telescope pointing resolution. Tracks were required to have hits in all six telescope planes and $ \chi^{2}/NDF \le 1.6 $. Events with more than one reconstructed track were rejected, to minimize reconstruction ambiguities stemming from, for instance, two separate pions traversing the telescope within the same timestamp. In addition, the projection on the DUT plane of the hits in each telescope plane were requested to be within a  window of $\SI{1.25}\times\SI{1.25} {\milli\meter^2}$ around the pixel with the largest number of entries. Tracks with hits outside this window in any of the planes were discarded, restricting the analyzed sample to tracks that crossed the DUT with a relatively small angle. For the tracks fulfilling the above selection, the telescope pointing resolution on the DUT plane was estimated to be approximately \SI{10}{\um}~\cite{mateus_thesis}. 

\subsection{Acquisition and Processing of the Signals from the DUT}\label{DUTsignals}
\begin{figure}[!htb]
\centering
\includegraphics[width=.99\textwidth,trim=10 10 10 10]{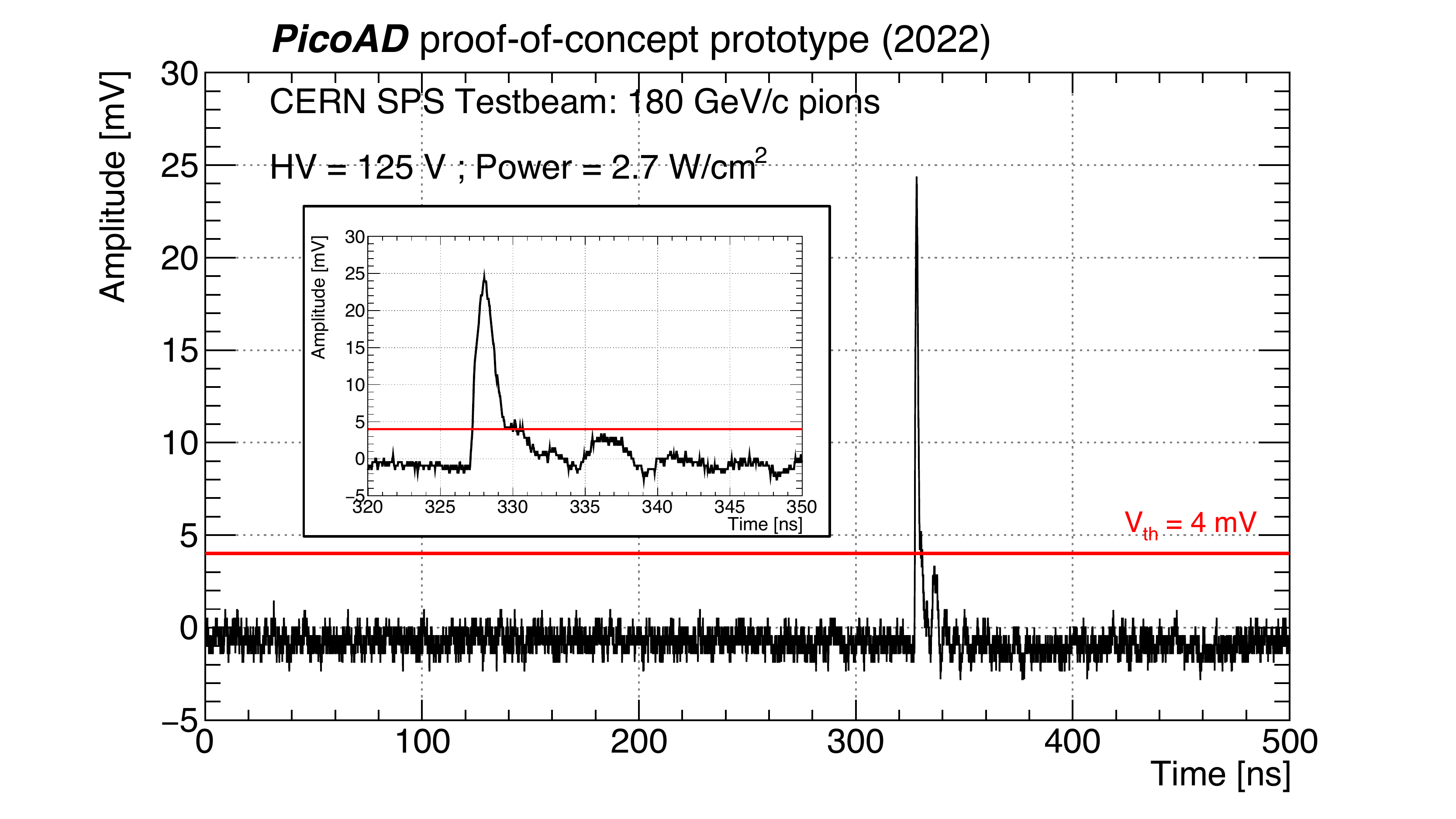}
\caption{\label{fig:waveform} Example of a small-amplitude waveform acquired for pixel OA0 with the DUT operated at preamplifier bias current $ \ipreamp = \SI{150}{\micro\ampere} $ and bias voltage $HV=\SI{125}{\volt}$. The time region below $\SI{250}{\nano\second}$ is the portion of the waveform used to extract the voltage noise $\sigma_{V}$. 
The insert  shows the same signal in a magnified time scale.
The full red lines show the discrimination threshold $V_{\mathrm{\it th}}$ used in the analysis.
}
\end{figure}

The  waveforms acquired by the oscilloscopes were recorded and used in the analysis of the selected data. For each DUT pixel, the signal was delayed to the second half of the 500 ns waveform time window acquired by the oscilloscopes, so that the first half of the waveform could be used to determine the voltage noise $\sigma_V$ at the output of the analog front-end in each given event. Figure~\ref{fig:waveform} shows a typical small-amplitude waveform, acquired for pixel OA0 with the DUT operated at preamplifier bias current $ \ipreamp = \SI{150}{\micro\ampere} $ and bias voltage $HV=\SI{125}{\volt}$.

Since  the DUT signal amplitudes were sampled at finite time intervals ($\SI{25}{\pico\second}$ for the first oscilloscope and $\SI{50}{\pico\second}$ for the second), a linear interpolation between samplings was performed to obtain the signal characteristics.  Only signals exceeding a discrimination threshold $ V_{\it{th}} = \SI{4}{\mV}$ in any of the DUT pixels were considered. In all  runs, the $\SI{4}{\mV}$ threshold was always  larger than $6\times\sigma_V$. 
The Time-Of-Arrival (TOA)  was defined as the time value at which the interpolated signal passed the threshold of $\SI{4}{\mV}$. 

Cross talk between the four analog channels\footnote{The cross talk could be ascribed to two possible causes: a feedback path passing through the ground of the board, or  noise induced by the driver pulse propagating in the power supply. A new ASIC prototype with separate power supplies for the pre-amplifier and for the driver stage has been recently produced. Preliminary measurements show that this new prototype is not affected by cross talk.} in events with large charge deposition was observed  for the ASIC version without gain  \cite{Iacobucci:2021ukp} that was produced with the same masks of the PicoAD proof-of-concept discussed here. 
As expected, this cross talk is present also in this PicoAD prototype. The insert in Figure~\ref{fig:waveform} shows the  delayed cross-talk, clearly visible after the real signal.
The same strategy of~\cite{Iacobucci:2021ukp} was thus adopted to suppress its impact: no additional selection was applied for the PicoAD prototype efficiency measurement, as cross talk may only appear for efficient events, while for the time resolution measurement only the pixel with the largest signal amplitude among the four pixels acquired was considered in each event.

\subsection{Data Samples}

Data were acquired at three  working points with amplifier bias current $\ipreamp = $  \SI{20}, \SI{50} and \SI{150}{\micro\ampere}, corresponding to a power density $P_{\mathrm {\it density}} =$ 0.4, 0.9 and 2.7 W/cm$^2$, respectively, to study the response of the frontend electronics. For the working point with $\ipreamp = $ \SI{150}{\micro\ampere}, a high voltage scan was also performed by acquiring two additional datasets with $HV$ = 105 and 115 V. 
The number of triggered events in each data sample
is reported in Table~\ref{tab:datasets}, which contains also the average number of  telescope tracks retained for each of the four DUT pixels after  the selection described in Section~\ref{selection}.
The large reduction in the number of selected telescope tracks with respect to recorded events is due partly to the trigger region-of-interest, that as described in Section~\ref{sec:setup} is much larger than the DUT area, and partly to the stringent telescope-track quality selection described in~Section~\ref{selection}.

\begin{table}[!htb]
\centering
\renewcommand{\arraystretch}{1.}
\begin{tabular}{|cc|cc|}
\cline{1-4}
\multicolumn{2}{|c|}{Working point} & \multirow{2}{*}{\makecell{Total events \\ recorded}} & \multirow{2}{*}{\makecell{Selected tracks \\ per DUT pixel}} \\ 
 $\ipreamp [\si{\micro\ampere}] $ & $HV [\si{\volt}]$ &  & \\
\cline{1-4}
150 & 125  &  888k & 67k  \\
150 & 115  &  224k & 16k  \\
150 & 105  &  152k & 12k  \\
 50 & 125  &  194k & 25k  \\
 20 & 125  & 795k & 64k  \\
\cline{1-4}
\end{tabular}
\caption{Number of recorded events and of the average selected telescope tracks pointing to each DUT pixel for the five  working points studied.
}
\label{tab:datasets} 
\end{table}

\subsection{DUT Signal Amplitudes}
\label{sec:radial}

To  inspect in detail the performance of this proof-of-concept PicoAD prototype and acquire information that could help improving the design of future  prototypes,
variations of the DUT  performance within the pixel area were investigated using the telescope pointing capabilities. With $r$ being the distance of the center of the hexagonal pixel from the point of intersection between the track reconstructed by the telescope and the DUT plane, the pixel area was divided into five radial regions containing  a similar number of telescope tracks: $r\le\SI{25}{\um}$,  $\SI{25}{\um}<r\le\SI{38}{\um}$, $\SI{38}{\um}<r\le\SI{48}{\um}$, $\SI{48}{\um}<r\le\SI{55}{\um}$ and $r>\SI{55}{\um}$. 

\begin{figure}[!htb]
\centering %
\includegraphics[width=.32\textwidth,trim=0 0 0 0, clip]{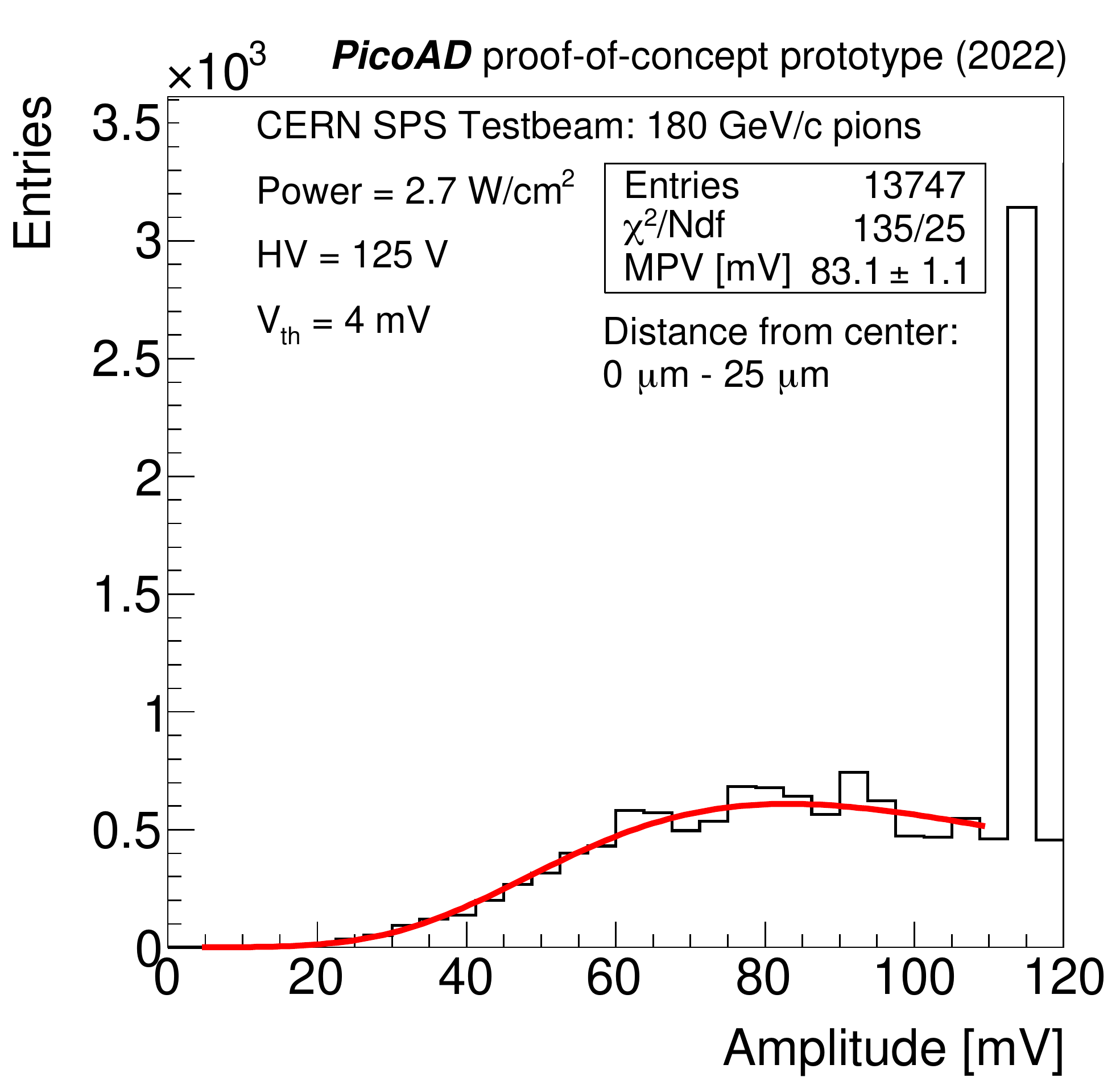}
\includegraphics[width=.32\textwidth,trim=0 0 0 0, clip]{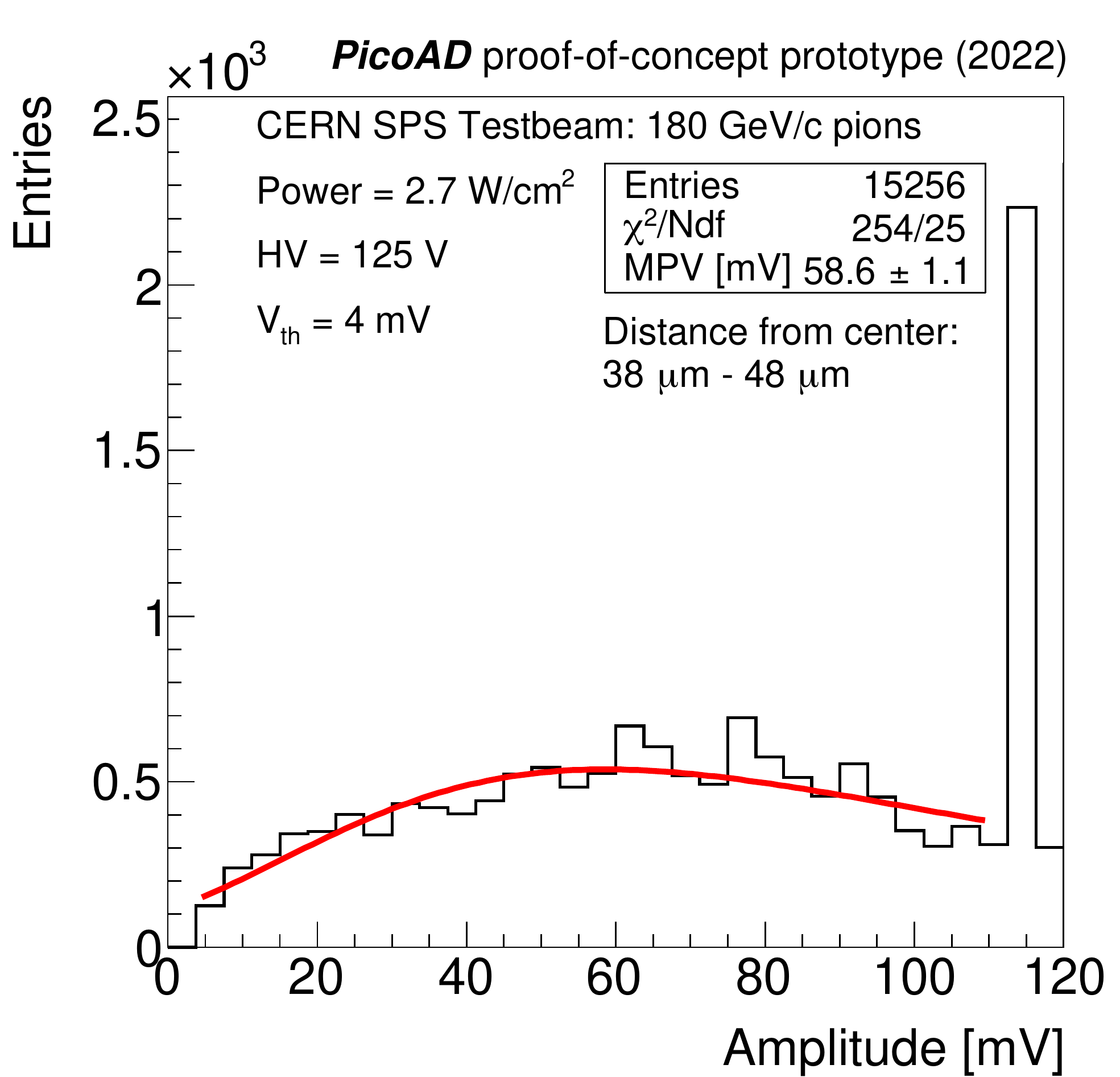}
\includegraphics[width=.32\textwidth,trim=0 0 0 0, clip]{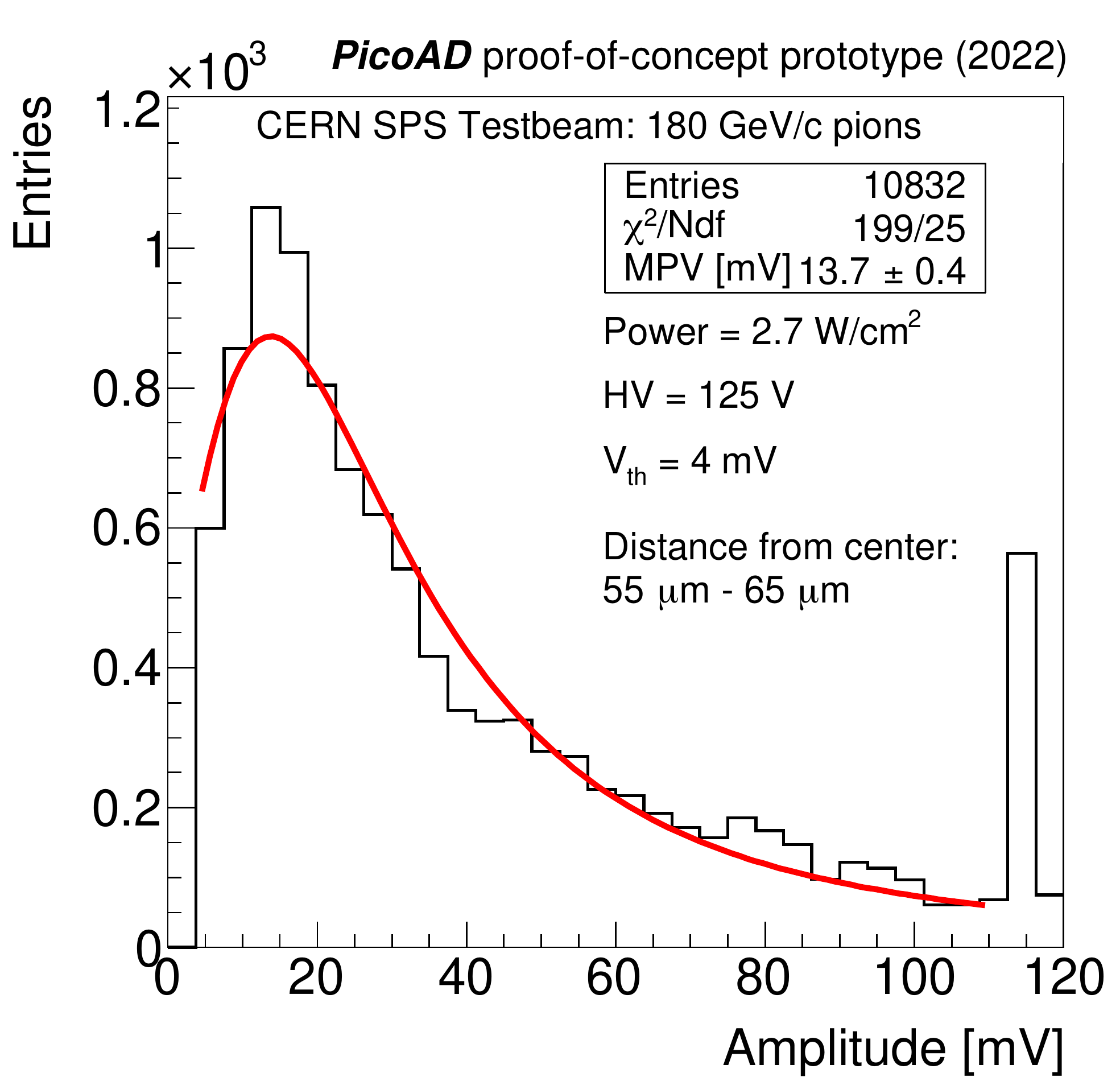}
\caption{\label{fig:amplitudes} 
Amplitude distribution of signals  from pixel OA0 for a power density of 2.7 W/cm$^2$ and $HV=\SI{125}{\volt}$ for telescope tracks at distances within $\SI{0}-\SI{25}{\micro\meter}$ (right), $\SI{38}-\SI{48}{\micro\meter}$ (center) and $\SI{55}-\SI{65}{\micro\meter}$ (right) from the  pixel center. The red lines show the fitted Landau probability distribution function performed in each distance range to obtain the most probable value (MPV) of the amplitude. The saturated amplitudes near $\SI{120}{\mV}$ (not fitted)are produced by the oscilloscope setting.
}
\end{figure}

\begin{figure}[!htb]
\centering %
\includegraphics[width=.49\textwidth,trim=0 0 0 0, clip]{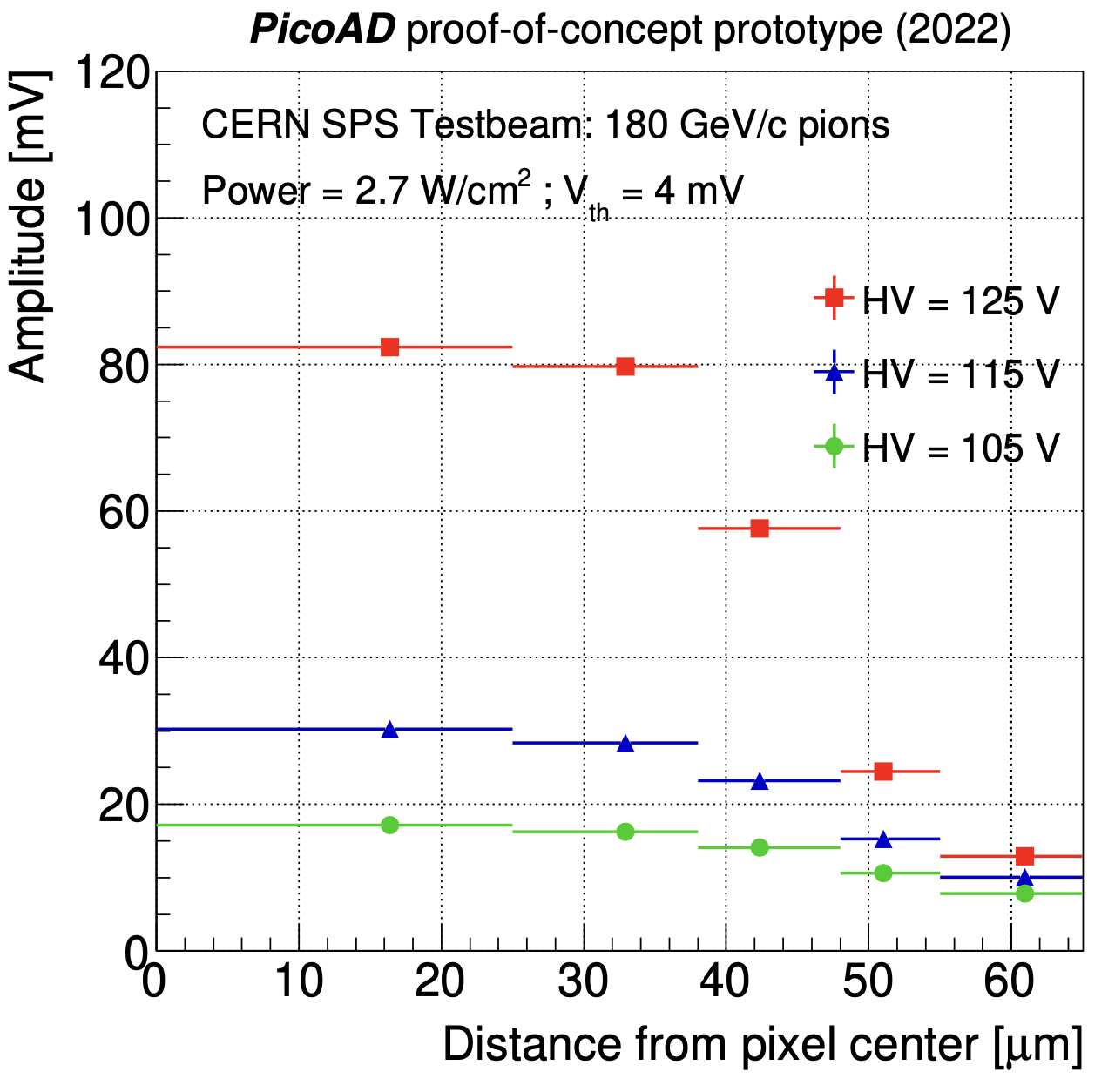}
\caption{\label{fig:MPV_amplitude} 
Most probable value of the signal amplitude recorded by pixel OA0 of the DUT as a function of the distance from the pixel center. The data refer to the  three values of sensor bias voltage acquired and at a power density of 2.7 W/cm$^2$. 
The  statistical uncertainty from the Landau probability distribution function fit is  within the marker size. In each  bin, the data points are positioned at the mean value of the distance from the center of the pixel. 
}
\end{figure}

The variation of the signal amplitude in these five radial regions  was studied for pixel OA0 of the DUT for different values of the sensor bias voltage.
For this purpose, the signal amplitude distributions in each  bin in distance $r$ and $HV$ value  were fitted with a Landau probability distribution function.
Figure~\ref{fig:amplitudes} shows three examples of the fitted distributions.
The most probable amplitude value  obtained by the fit in each distance bin was then used to quantify the amplitude variation between the center of the pixel and its edge. 
The result is shown in Figure~\ref{fig:MPV_amplitude}  for the three values of the sensor bias voltage considered. 
As expected, the signal amplitude  depends on the sensor bias voltage; this is explained by the increasing gain with sensor bias voltage~\cite{picoad_gain}, that for minimum-ionizing particles  produces an average of 5k, 8k and 20k electrons at 105, 115 and 125 $V$, respectively.
The sizeable drop in amplitude towards the edge of the pixel\footnote{
The pixel edge is at a distance from the pixel center   
between 56 µm (center of the hexagon side) and 65 µm (corner of three pixels).} 
is due partly to charge sharing between adjacent pixels, and partly to the lower electric field under the p-stop structure 
that separates electrically the  pixels~\cite{picoad_gain}.
It should be noted that the telescope spatial resolution of approximately 10 µm implies that migrations between adjacent bins of the distance $r$ from the pixel center are not negligible. These migrations were not unfolded in the results shown in Figure~\ref{fig:MPV_amplitude} as well as in the other results as a function of $r$ that are reported below.

%% file: Efficiency.tex
\section{Efficiency Measurement}
\label{sec:efficiency}

The detection efficiency was defined as the fraction of  selected telescope tracks  that led to a recorded signal above the discrimination threshold in one of the DUT pixels. Only tracks crossing the area of the four instrumented pixels were considered, with a tolerance of $\SI{10}{\um}$ beyond the pixel edges, which corresponds to one standard deviation of the telescope pointing resolution. 


\begin{figure}[!htb]
\centering
\includegraphics[width=.49\textwidth,trim=0 0 0 0]{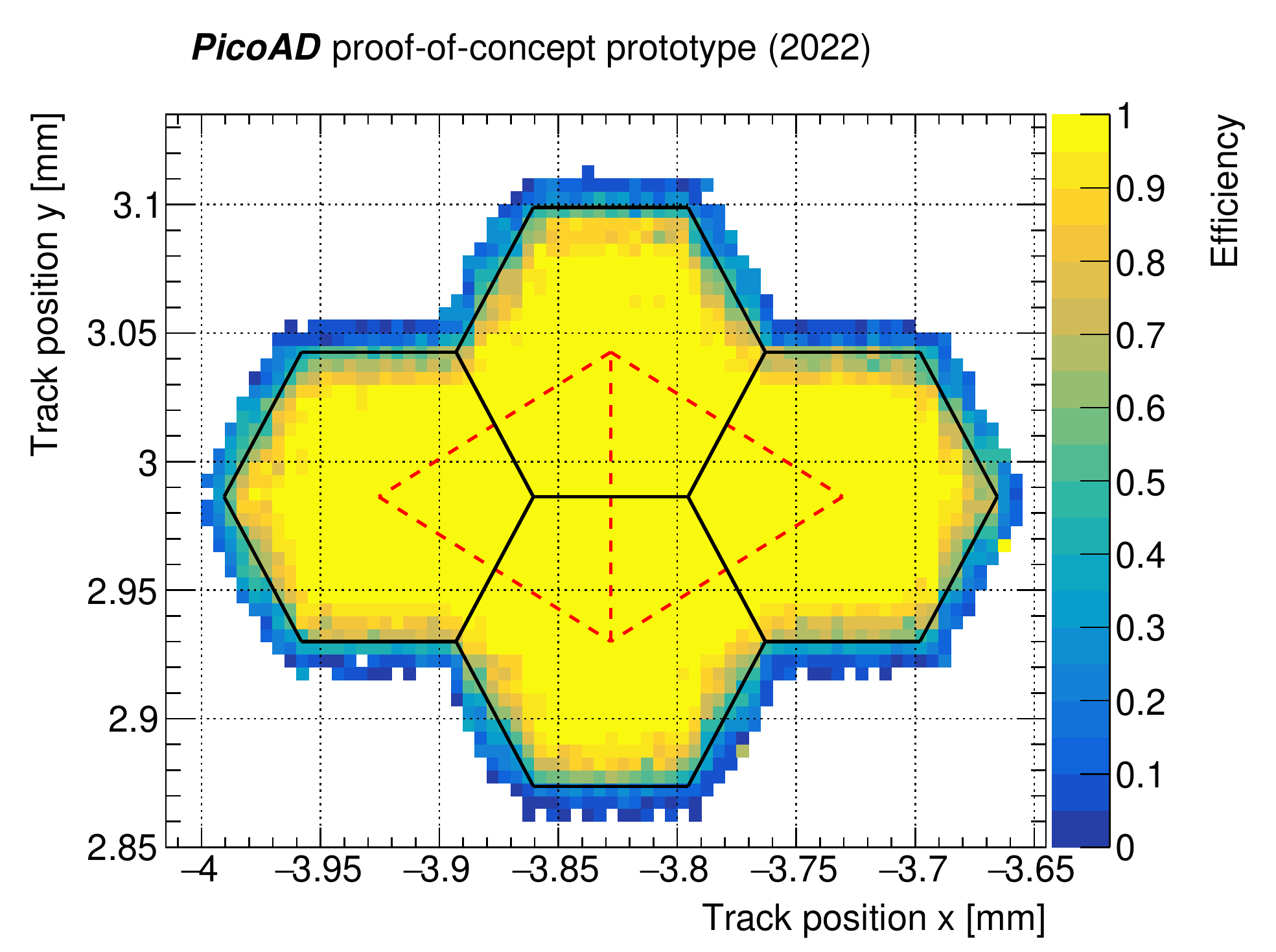}
\includegraphics[width=.49\textwidth,trim=0 0 0 0]{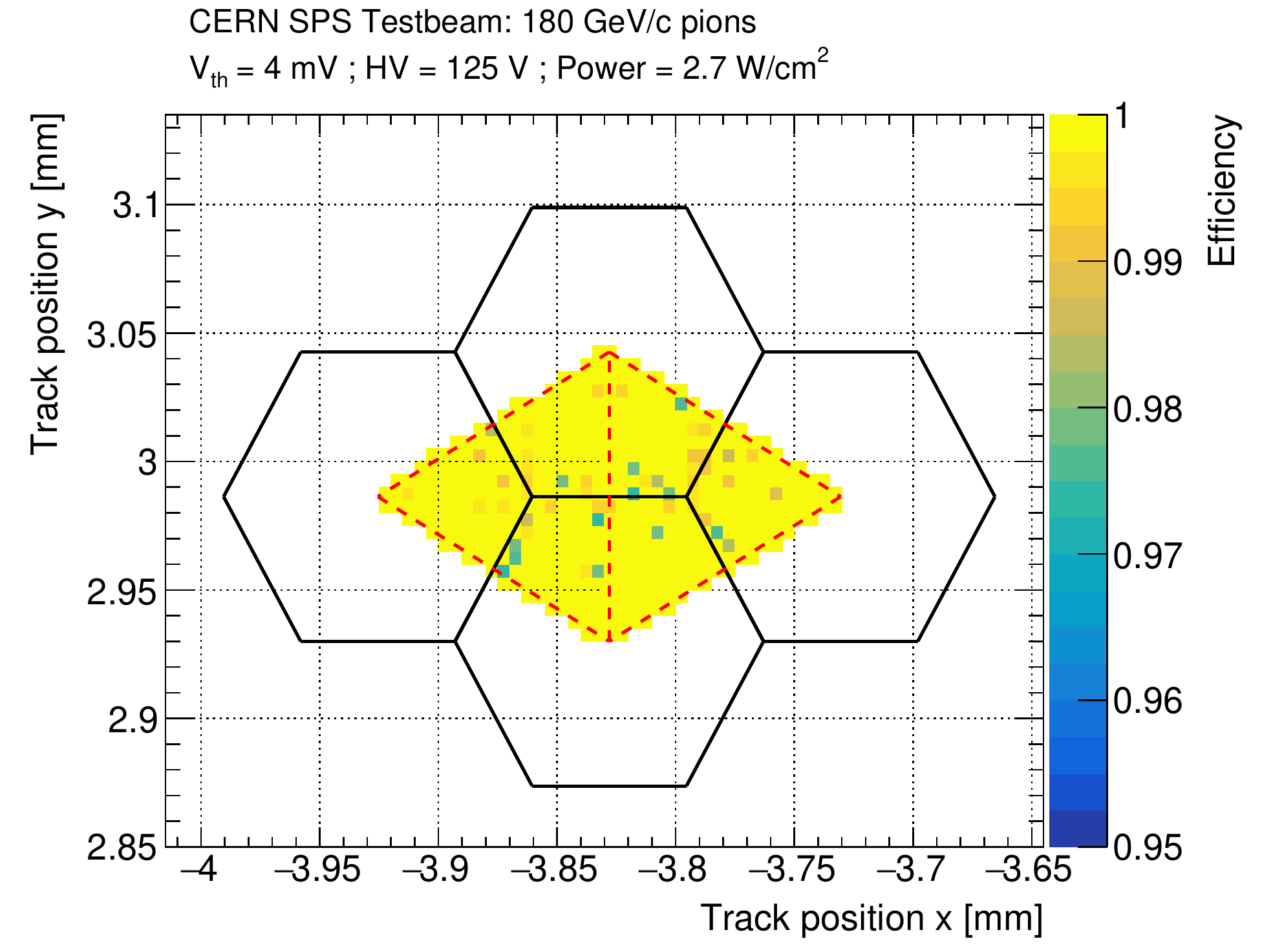}
\caption{\label{fig:effmap} Efficiency measured for the DUT operated at a power density of 2.7 W/cm$^2$
 for a discrimination threshold $V_{\mathrm {\it th}}=\SI{4}{\mV}$. The black lines indicate the pixel edges.  The left panel shows the results for the entire surface of the four analog pixels. The apparent efficiency degradation around the external edges is a consequence of the FEI4 telescope pointing resolution. The right panel illustrates the efficiency in the two triangular areas delimited by the centers of the three {\it left}-most and {\it right}-most pixels, which is unaffected by the telescope resolution and is therefore used throughout this study. To highlight the small  inefficiencies, measured mostly in the inter-pixel regions, the color scale in the right panel starts from 95\%.
}
\end{figure}

The left panel of Figure~\ref{fig:effmap} illustrates the efficiency measured in the whole area corresponding to the four analog pixels when  the DUT was operated at $P_{\mathrm {\it density}} =$ 2.7 W/cm$^2$ (or equivalently $\ipreamp = \SI{150}{\uA}$) and $HV=\SI{125}{\volt}$. The apparent degradation of the detection efficiency observed around the external edges of the four pixels can be attributed to the telescope pointing resolution of 10 µm, since no efficiency degradation is visible in the five  internal inter-pixel boundaries. To overcome the bias from the telescope resolution, the efficiency measurement was then restricted to the two triangular areas shown in Figure~\ref{fig:effmap} right, defined by the centers of the three left-most (OA0, OA2, OA3) and right-most (OA0, OA2, OA1) pixels. Each triangle covers exactly one sixth of the area of the three hexagonal pixels involved, and contains the boundaries between two pixels and the three-pixel corners 
in the same proportion as in a pixel.
They provide therefore a good representation of the efficiency behavior across an entire pixel. With few exceptions, the  bins inside the two triangles corresponding to slightly lower measured efficiencies  tend to accumulate near the pixel borders, which can be explained by the lower signal amplitudes in these regions displayed in Figure~\ref{fig:MPV_amplitude}.

\begin{figure}[!htb]
\centering 
\includegraphics[width=.49\textwidth,trim=0 0 0 0]{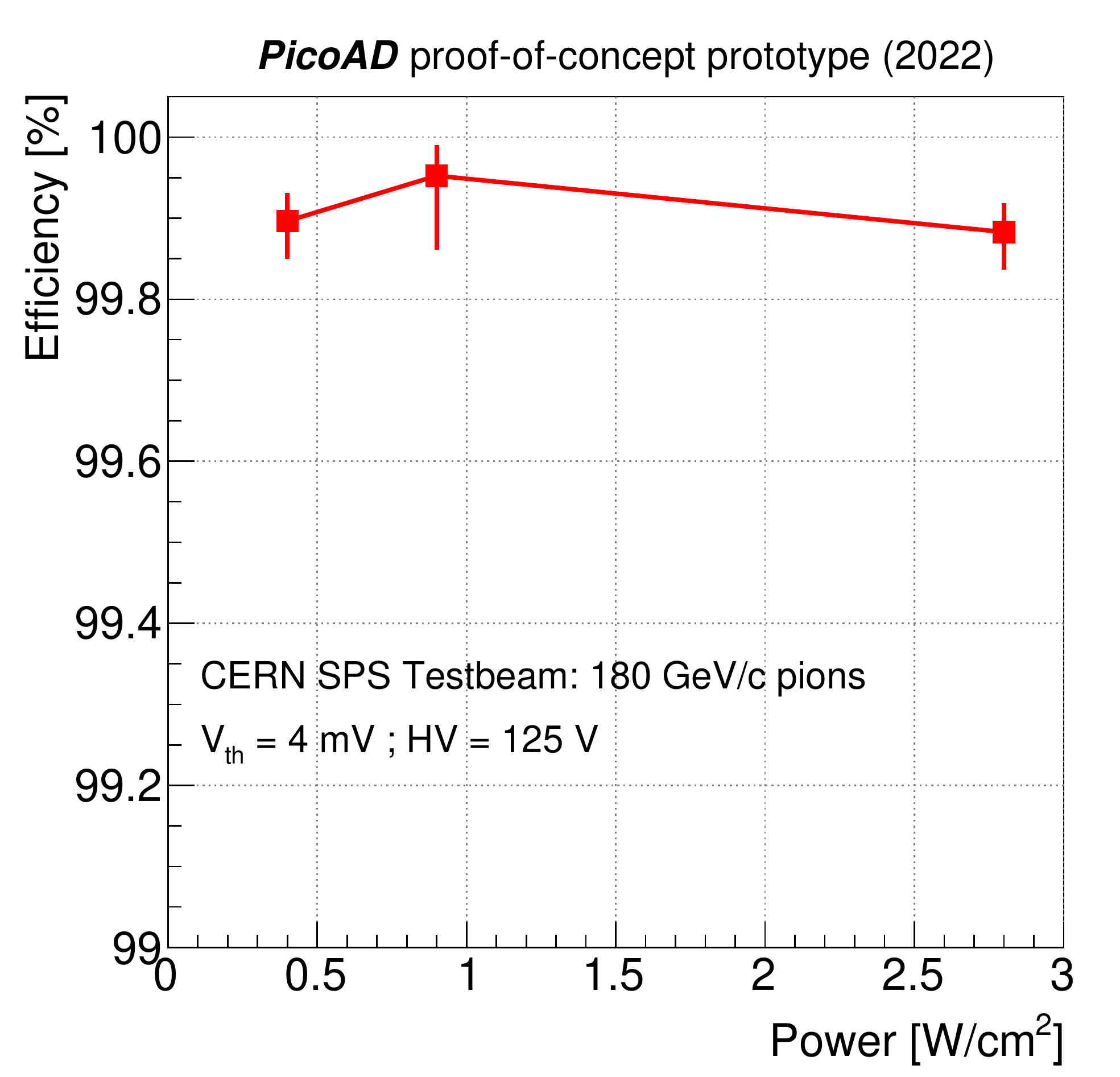}
\includegraphics[width=.49\textwidth,trim=0 0 0 0]{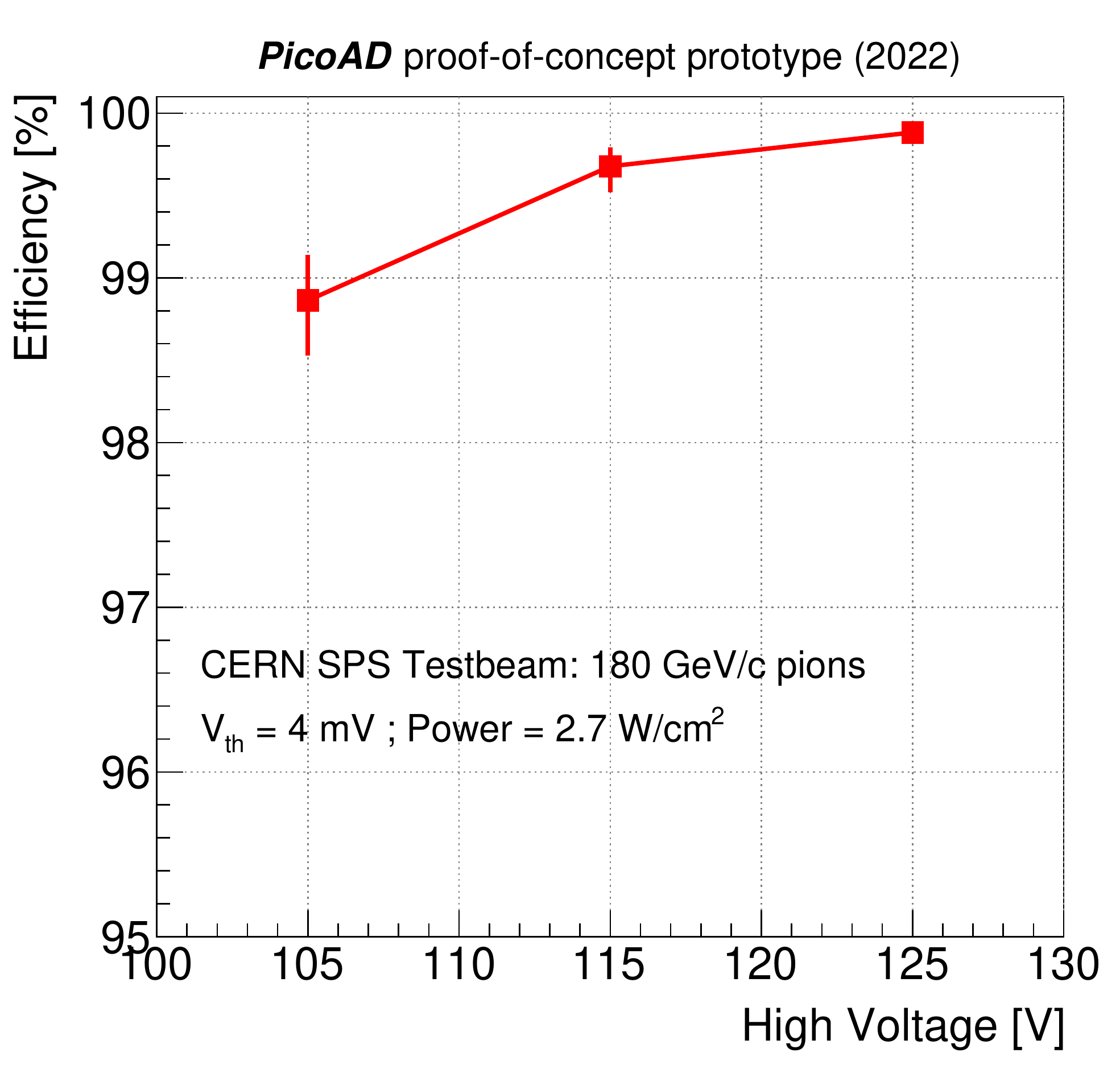}
\caption{\label{fig:eff_ipream_HV_pscan} Detection efficiency measured within the two triangular areas shown in~Figure~\ref{fig:effmap} with $V_{\mathrm{\it th}} = \SI{4}{\mV}$. The left panel shows  the  efficiency  measured at $HV$ = 125 V for three power density values considered, while the right panel shows the  efficiency  measured for   three values of $HV$ at a power density of 2.7 W/cm$^2$ (corresponding to $ \ipreamp = \SI{150}{\micro\ampere}$). 
}
\end{figure}

Table~\ref{tab:eff_ipream_HV_pscan} summarizes the efficiency measured in the left and  right triangles shown in Figure~\ref{fig:effmap} for the five working points at which data were acquired. For each working point, the average of the two measurements is reported in the last column of Table~\ref{tab:eff_ipream_HV_pscan} and in Figure~\ref{fig:eff_ipream_HV_pscan}.
In the case in which  the sensor is operated at a bias voltage $ HV = \SI{125}{\volt} $, the efficiencies for the
three values of power consumption considered are observed to be compatible with 99.9\% within uncertainties.
At the highest power consumption measured
$P_{\mathrm {\it density}} =$ 2.7 W/cm$^2$, 
a small decrease in detection efficiency is observed at  $HV$ = 105 and 115 V, consistently with the expected reduction of the sensor gain at these sensor bias voltages.

\begin{table}[!htb]
\centering 
\renewcommand{\arraystretch}{1.4}
\begin{tabular}{cccccc|}
\cline{1-6}
\multicolumn{1}{|c|}{ \multirow{2}{*}{$\ipreamp$ [\si{\uA}]} } & \multicolumn{1}{c|}{\multirow{2}{*}{$P_{\mathrm {\it density}}$ [W/cm$^2$]} } & \multicolumn{1}{c|}{\multirow{2}{*}{ $HV$ [\si{V}]} } & \multicolumn{3}{c|}{Efficiency [\%] }  \\ 
\cline{4-6}
\multicolumn{1}{|c|}{ }  & \multicolumn{1}{c|}{ } & \multicolumn{1}{c|}{ } & \multicolumn{1}{c|}{~~{\it Left} triangle~~ } & \multicolumn{1}{c|}{~{\it Right} triangle~ } & \multicolumn{1}{c|}{~~~~~Average~~~~~ } \\ 
\cline{1-6}
\multicolumn{1}{|c|}{20}  & \multicolumn{1}{c|}{0.4} & \multicolumn{1}{c|}{125} & \multicolumn{1}{c|}{$ 99.92_{-0.04}^{+0.03} $} & \multicolumn{1}{c|}{$  99.87_{-0.04}^{+0.05} $} & \multicolumn{1}{c|}{$ 99.90_{-0.05}^{+0.04} $} \\ 
\multicolumn{1}{|c|}{50}  & \multicolumn{1}{c|}{0.9} & \multicolumn{1}{c|}{125} & \multicolumn{1}{c|}{$ 99.95_{-0.09}^{+0.04} $} & \multicolumn{1}{c|}{$ 99.95_{-0.09}^{+0.04} $} & \multicolumn{1}{c|}{$ 99.95_{-0.09}^{+0.04} $} \\ 
\multicolumn{1}{|c|}{150} & \multicolumn{1}{c|}{2.7} & \multicolumn{1}{c|}{125} & \multicolumn{1}{c|}{$ 99.87_{-0.05}^{+0.04} $} & \multicolumn{1}{c|}{$ 99.89_{-0.05}^{+0.03} $} & \multicolumn{1}{c|}{$ 99.88_{-0.05}^{+0.04} $}  \\ 
\multicolumn{1}{|c|}{150} & \multicolumn{1}{c|}{2.7} & \multicolumn{1}{c|}{105} & \multicolumn{1}{c|}{$ 99.00_{-0.03}^{+0.03} $} & \multicolumn{1}{c|}{$  98.75_{-0.04}^{+0.03} $} & \multicolumn{1}{c|}{$ 98.86_{-0.03}^{+0.03} $} \\ 
\multicolumn{1}{|c|}{150} & \multicolumn{1}{c|}{2.7} & \multicolumn{1}{c|}{115} & \multicolumn{1}{c|}{$ 99.73_{-0.15}^{+0.11} $} & \multicolumn{1}{c|}{$ 99.62_{-0.17}^{+0.13} $} & \multicolumn{1}{c|}{$ 99.68_{-0.16}^{+0.12} $} \\ 
\cline{1-6}
\end{tabular}
\caption{Detection efficiency measured for the PicoAD proof-of-concept prototype at different $\ipreamp$ and $ HV $ values. The efficiencies are separately extracted in the two triangular areas connecting the pixel centers of the three {\it left}-most (OA0, OA2, OA13) and {\it right}-most (OA0, OA2, OA3) pixels shown in Figure~\ref{fig:effmap}. The average efficiencies in the two triangles are also 
reported.
Uncertainties are statistical only.
}
\label{tab:eff_ipream_HV_pscan}
\end{table}

\begin{figure}[!htb]
\centering 
\includegraphics[width=.75\textwidth,trim=0 0 0 0]{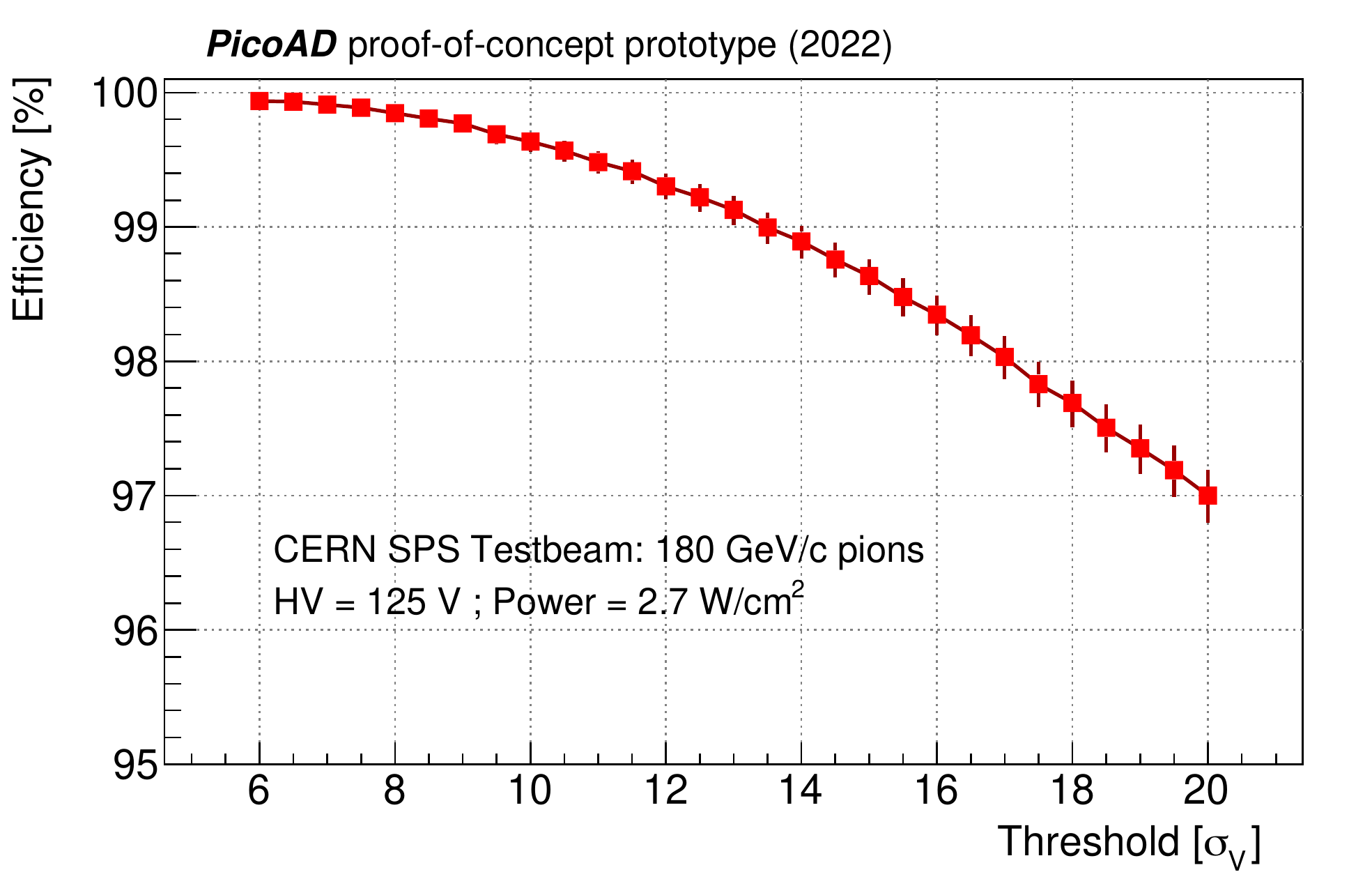}
\caption{\label{fig:effthrscan} Detection efficiency measured within the two triangular areas shown in~Figure~\ref{fig:effmap},  as a function of the discrimination threshold in multiples of the the voltage noise $\sigma_V$. The data refer to   $ HV = \SI{125}{\volt} $ and  $P_{\mathrm {\it density}} =$ 2.7 W/cm$^2$.
The 4 mV threshold used throughout this paper corresponds  approximately to 7.5 times  the voltage noise $\sigma_V$}
\end{figure}

Figure~\ref{fig:effthrscan} illustrates the results of a scan over the discrimination threshold $ V_{\mathrm{\it th}}$, taken at integer multiples of the voltage noise $\sigma_V$, for $P_{\mathrm {\it density}} =$ 2.7 W/cm$^2$ and $HV=\SI{125}{\volt}$. The efficiency measured in the two triangles drops below $99\%$ only for threshold values above $13\times\sigma_V$, demonstrating the capability of operating the ASIC with very high detection efficiency even at relatively high thresholds. 

\vspace{30pt}

\subsection{Efficiencies in  Different Regions of the Pixel}

\begin{figure}[!htb]
\centering 
\includegraphics[width=.49\textwidth,trim=0 0 0 0]{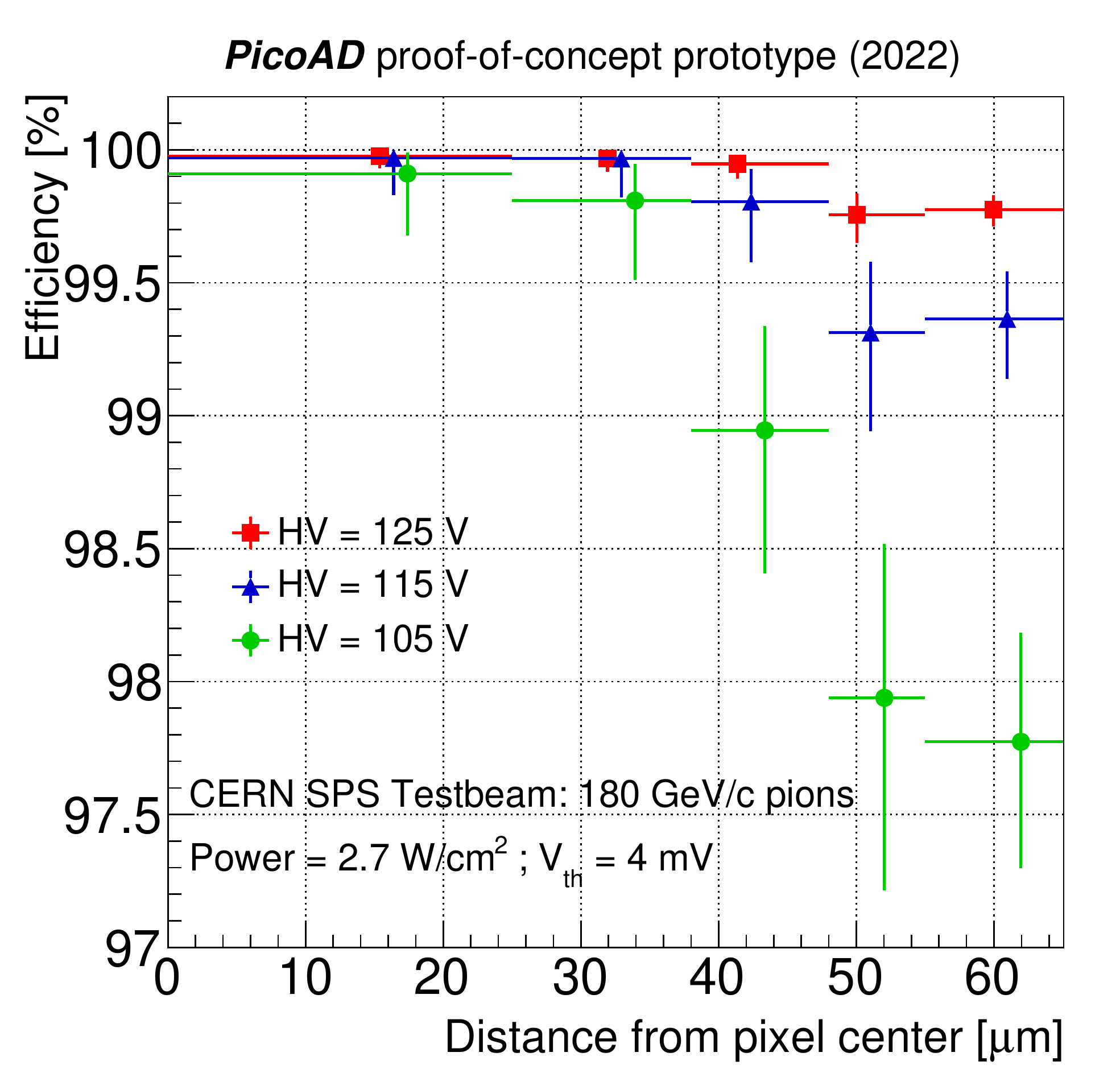}
\includegraphics[width=.49\textwidth,trim=0 0 0 0]{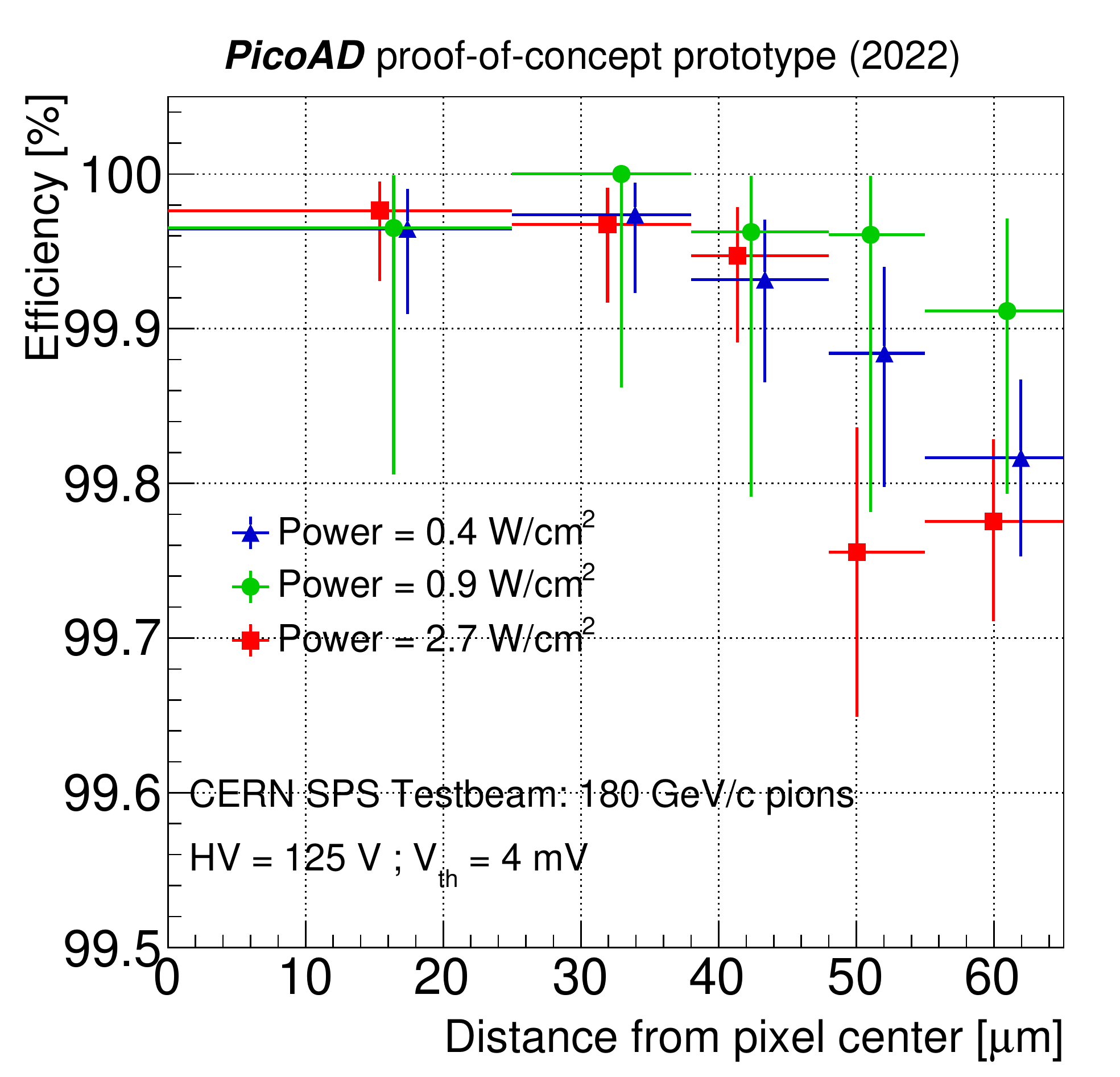}
\caption{\label{fig:eff_radius} Detection efficiency measured within the two triangular areas shown in~Figure~\ref{fig:effmap} with $V_{\mathrm{th}} = \SI{4}{\mV}$, shown as a function of the distance from the center of the pixels. The left panel shows the  efficiency measured for three values of $HV$ at a power density of 2.7 W/cm$^2$ (corresponding to $ \ipreamp = \SI{150}{\micro\ampere}$), while  the right panel shows the  efficiency  measured at $HV$ = 125 V for the three power density values. 
To improve readability, the data are slightly shifted in the horizontal scales.}
\end{figure}

The DUT efficiencies were measured as a function of the distance from the pixel center for the five  working points acquired.
Figure~\ref{fig:eff_radius} shows the efficiencies measured in the same five radial regions described in Section~\ref{sec:radial}. A drop in efficiency towards the edge of the pixel of approximately 2\% is observed  when the sensor bias decreases from 125 to 105 V. This can be explained by the reduced electric field present in this prototype under the p-stop inter-pixel structure, discussed in~\cite{picoad_gain}.
In the case of HV = 125 V, the efficiency drop at the edge of the pixel  is only 0.2\%, and compatible within uncertainties for the three power consumption values investigated here.  

%% file: Resolution.tex
\section{Time Resolution Measurement}
\label{sec:resolution}

The  time resolution of the PicoAD proof-of-concept prototype was measured by constructing a system of coincidences between the DUT and the two LGADs. The distribution of the difference in TOA ($\dtoa$), which gives the TOF between the detectors, was studied separately for the three  detector pairs, DUT-LGAD0, DUT-LGAD1 and LGAD0-LGAD1, and used to measure the corresponding standard deviation $\sigdtoa$. The result is a system of three equations, whose solution provides the DUT time resolution $\sigdut$ and the two LGAD time resolutions $\siglgzero$ and $\siglgone$:

\begin{equation}
\label{eq:system}
\begin{array}{lcl} 
\sigdtoa^2{_{({\it DUT,LGAD0})}} & = &  \sigdut^2+\siglgzero^2\\ 
\sigdtoa^2{_{({\it DUT,LGAD1})}} & = &  \sigdut^2+\siglgone^2\\ 
\sigdtoa^2{_{({\it LGAD0,LGAD1})}} & = &  \siglgzero^2+\siglgone^2\\ 
\end{array}
\end{equation}

The four  PicoAD analog pixels were considered individually. If a signal above threshold was recorded in more than one pixel, as in the case of  particles crossing inter-pixel regions and producing charge shared between adjacent pixels, the signal in the pixel with  largest amplitude was considered. 
The oscilloscopes produced a saturation of the amplitudes, to around 120 mV for pixel OA0 and  80 mV for the other three pixels.
Since in the rare case of two saturated signals it was not possible to select the pixel with highest amplitude,
events in which two pixels had  
amplitude above $\SI{75}{mV}$  were analyzed for both pixels. It was nonetheless checked that retaining or rejecting the small fraction of events with such ambiguity has no impact on the results of the time resolution measurement. 

\subsection{Time-Walk Correction}
\label{sec:twc}

\begin{figure}[!htb]
\centering
\includegraphics[width=.48\textwidth,trim=0 0 0 0]{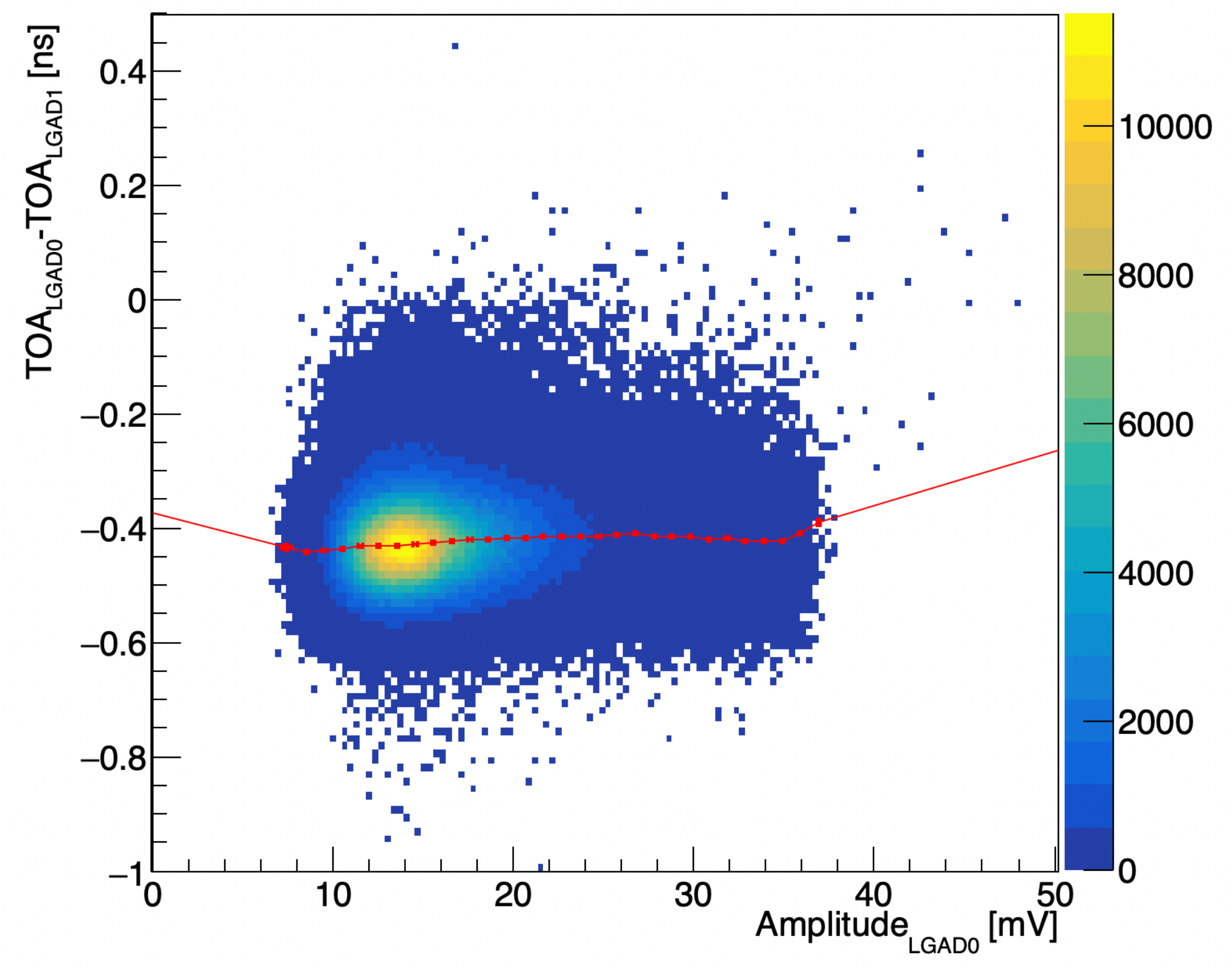}
~~
\includegraphics[width=.48\textwidth,trim=0 0 0 0]{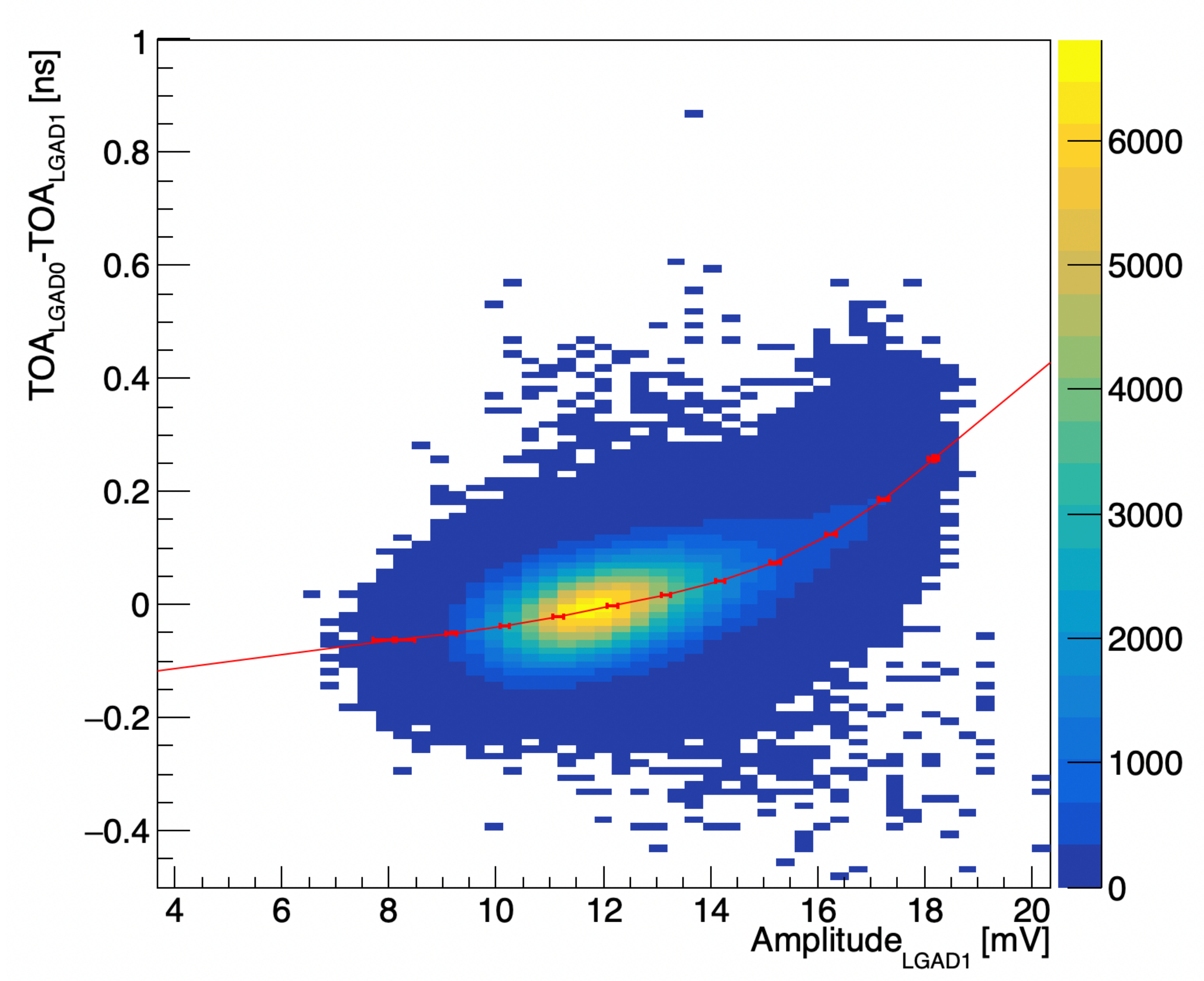}
\caption{\label{fig:twcorr_LGADS} Distributions of the $\dtoa$ difference between the LGAD0 and LGAD1 as a function of the  signal amplitudes measured by LGAD0 (left) and LGAD1 (right). The time-walk correction points (in red) were obtained by a Gaussian fit in each bin of the horizontal axis.  The red segments show the linear interpolation between the time-walk correction points used to correct the data. The $\dtoa$  contains an arbitrary offset that is irrelevant for the measurement of the time resolution.}
\end{figure}

\begin{figure}[!htb]
\centering
\includegraphics[width=.49\textwidth,trim=0 0 0 0]{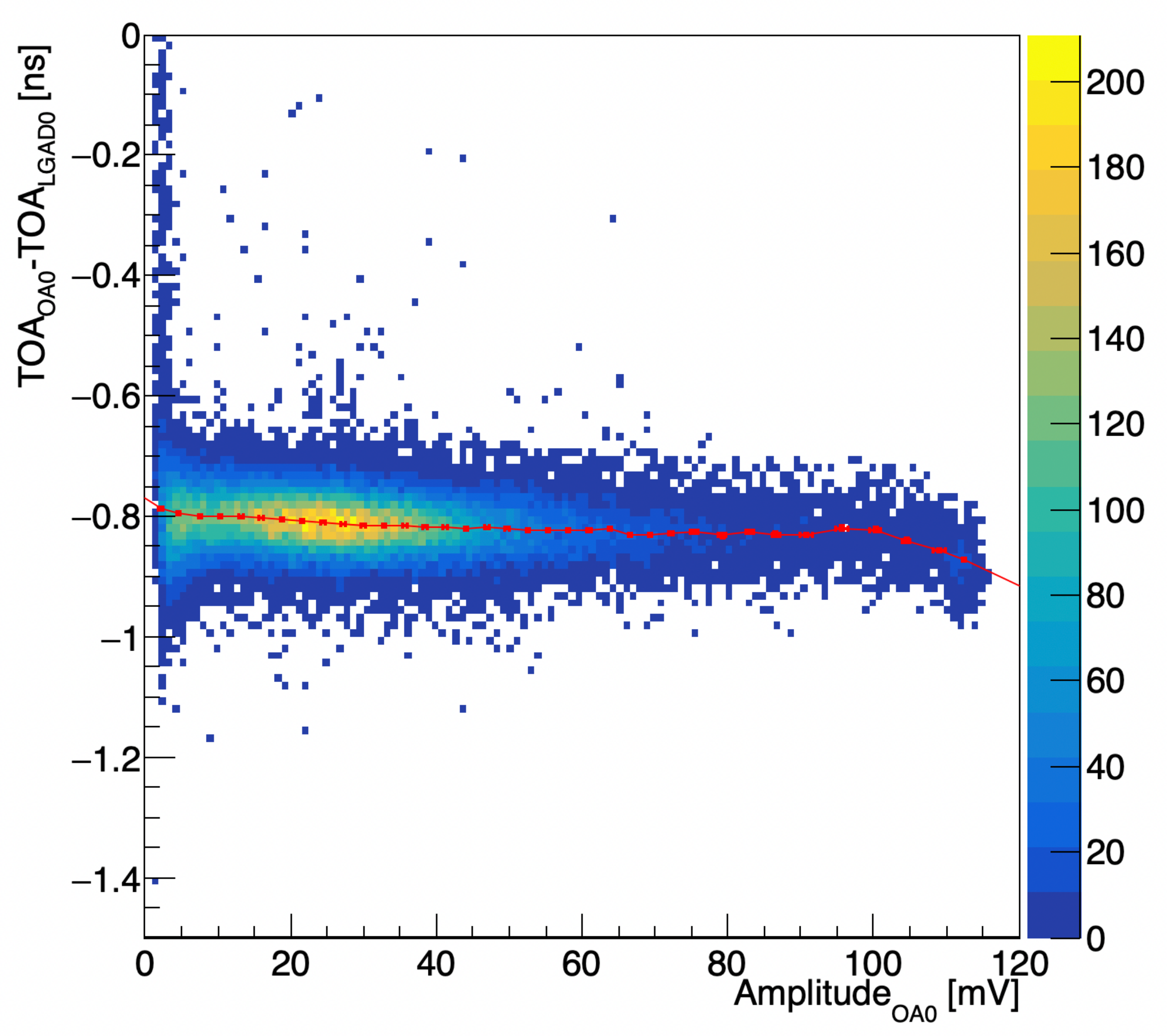}
\includegraphics[width=.49\textwidth,trim=0 0 0 0]{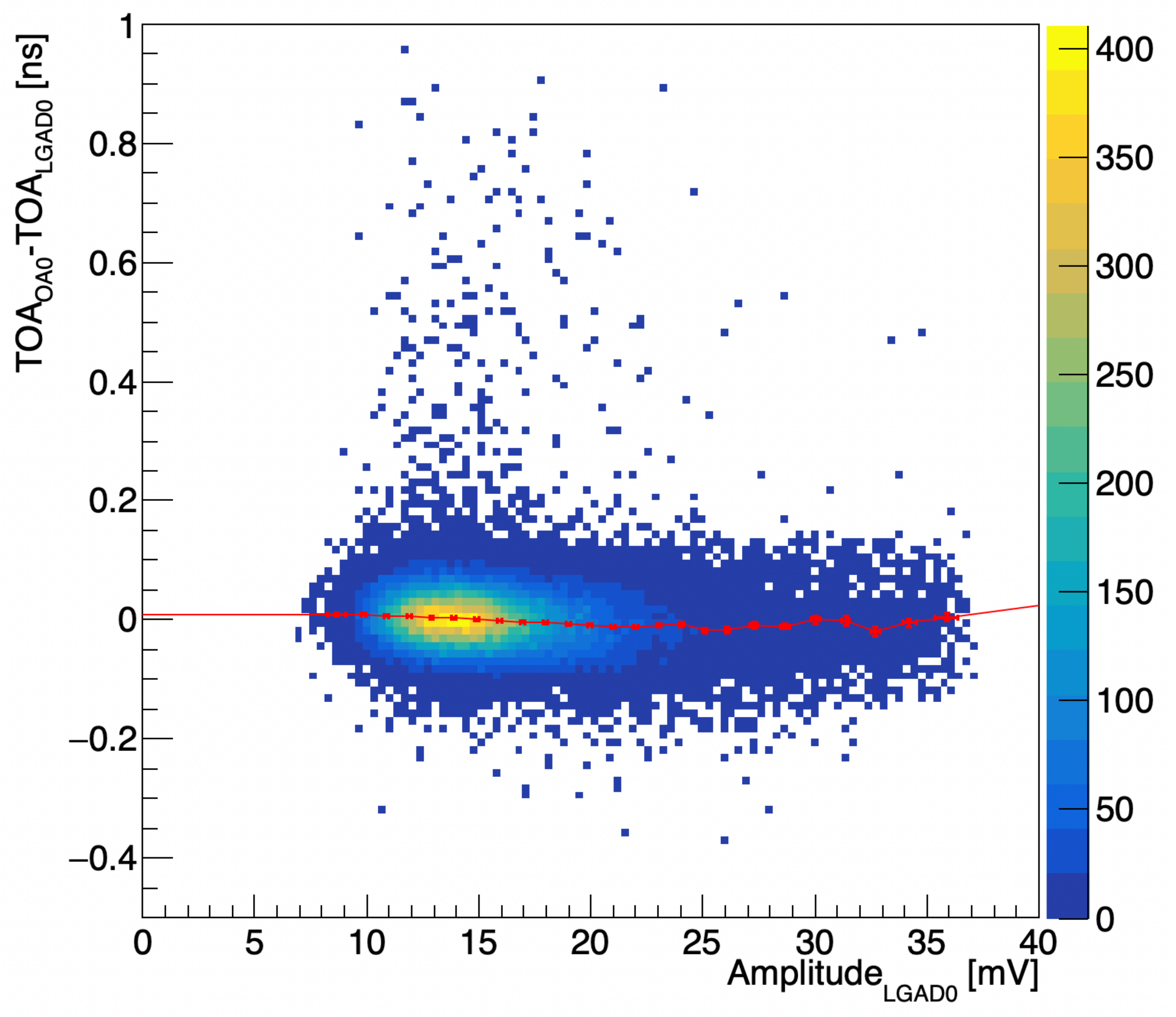}
\caption{\label{fig:twcorr_DUT} Distributions of the difference $\dtoa$  between the "CFD-like" signals in the pixel OA0 of the DUT and the signals acquired by LGAD0 as a function of the DUT amplitude (left) and LGAD0 amplitude (right). The DUT was operated at a power density of 2.7 W/cm$^2$ (corresponding to $ \ipreamp = \SI{150}{\micro\ampere}$) and $ HV = \SI{125}{\volt} $. The time-walk correction points (in red) were obtained by a Gaussian fit on each bin of the horizontal axis. The red segments show the linear interpolation between the time-walk correction points used to correct the data. The $\dtoa$ difference contains an arbitrary offset that is irrelevant for the measurement of the time resolution.}
\end{figure}

Different signals may cross the discrimination thresholds with different delays with respect to the instant when a ionizing particle crosses the sensor depending on the signal amplitude and risetime. This effect, usually referred to as time walk, leads to the  variation of the average of the difference $\dtoa$  as a function of the signal amplitude that is visible in the plots of Figures~\ref{fig:twcorr_LGADS} and ~\ref{fig:twcorr_DUT}. 

An event-by-event correction was extracted to mitigate the effect of time walk on the measured time resolution. For each of the bins of the horizontal axes of the plots in Figures~\ref{fig:twcorr_LGADS} and ~\ref{fig:twcorr_DUT}, the most probable value of $\dtoa$ was extracted via a Gaussian fit (red points in the figures), and associated to the average signal amplitude for that bin. A linear interpolation between adjacent bins (red lines in the figures) was then used to compute the correction factor to be applied to the $\dtoa$ value of each event. More precisely, a first correction was determined for the first detector in the pair associated to $\dtoa$, and then applied; subsequently, a second correction was determined for the second detector, and applied to obtain the final $\dtoa$ distribution corrected for time walk. 
It was verified that 
the order of the correction is irrelevant,
as expected for  two  independent detectors. 

While the time-walk correction for the LGADs was always computed using their corresponding signal amplitudes,  different approaches were studied for the PicoAD prototype. 
\begin{itemize}
\item A first time-walk correction was obtained relying solely on the DUT signal amplitudes, as adopted in~\cite{Iacobucci:2021ukp}.

\item
A second approach consisted in shifting by 200 ps (approximately half of the signal  risetime) the  waveform from the DUT pixel,  subtracting it from the original waveform, and taking as TOA the time at 25\% height of the produced waveform. This method, that  mimics at some level the working principle of a constant fraction discriminator (CFD) and is referred here as "CFD-like", 
avoids time walk between large and small amplitude signals, while  at the same time reduces the small frequency-noise components that are automatically suppressed when subtracting the original and time-shifted signals to generate the CFD-like signal. 
The small residual time walk was then corrected using the DUT amplitude of the new waveform and then that of the LGAD, as shown in Figure~\ref{fig:twcorr_DUT}.
\item Finally, a hybrid approach that used the amplitude to correct   for time walk signals with amplitudes smaller than 25 mV and the signal risetime between 10 and 20 mV for signals larger than 25 mV, was also utilized. This method gave results compatible within uncertainties with those obtained with the CFD-like method.
\end{itemize}

The CFD-like approach was adopted to obtain the results shown in this paper.

\subsection{Estimation of the Time Jitter Between the Two Oscilloscopes}\label{oscjitter}

\begin{figure}[!htb]
\centering %
\includegraphics[width=.49\textwidth,trim=0 0 0 0]{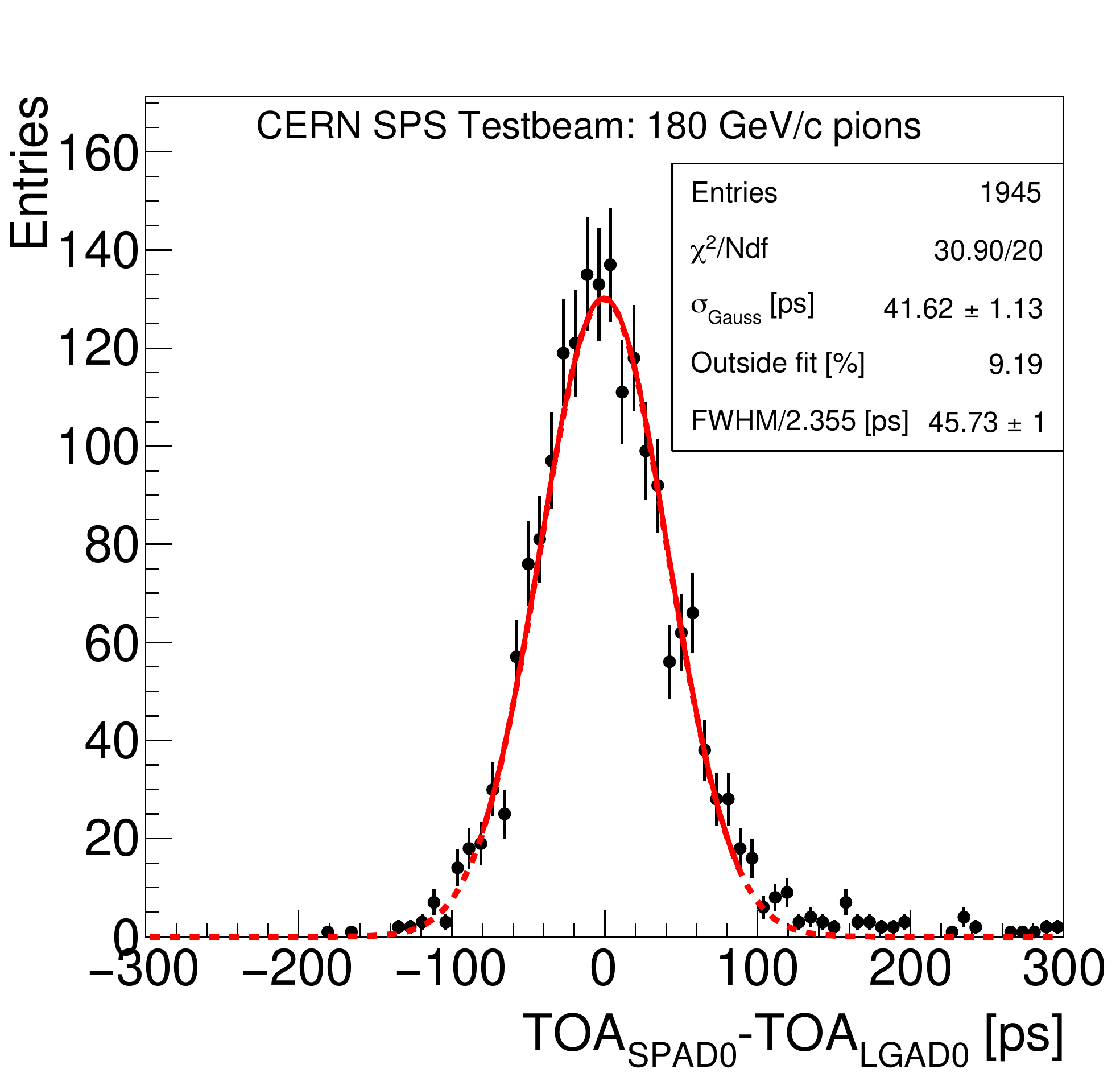}
\includegraphics[width=.49\textwidth,trim=0 0 0 0]{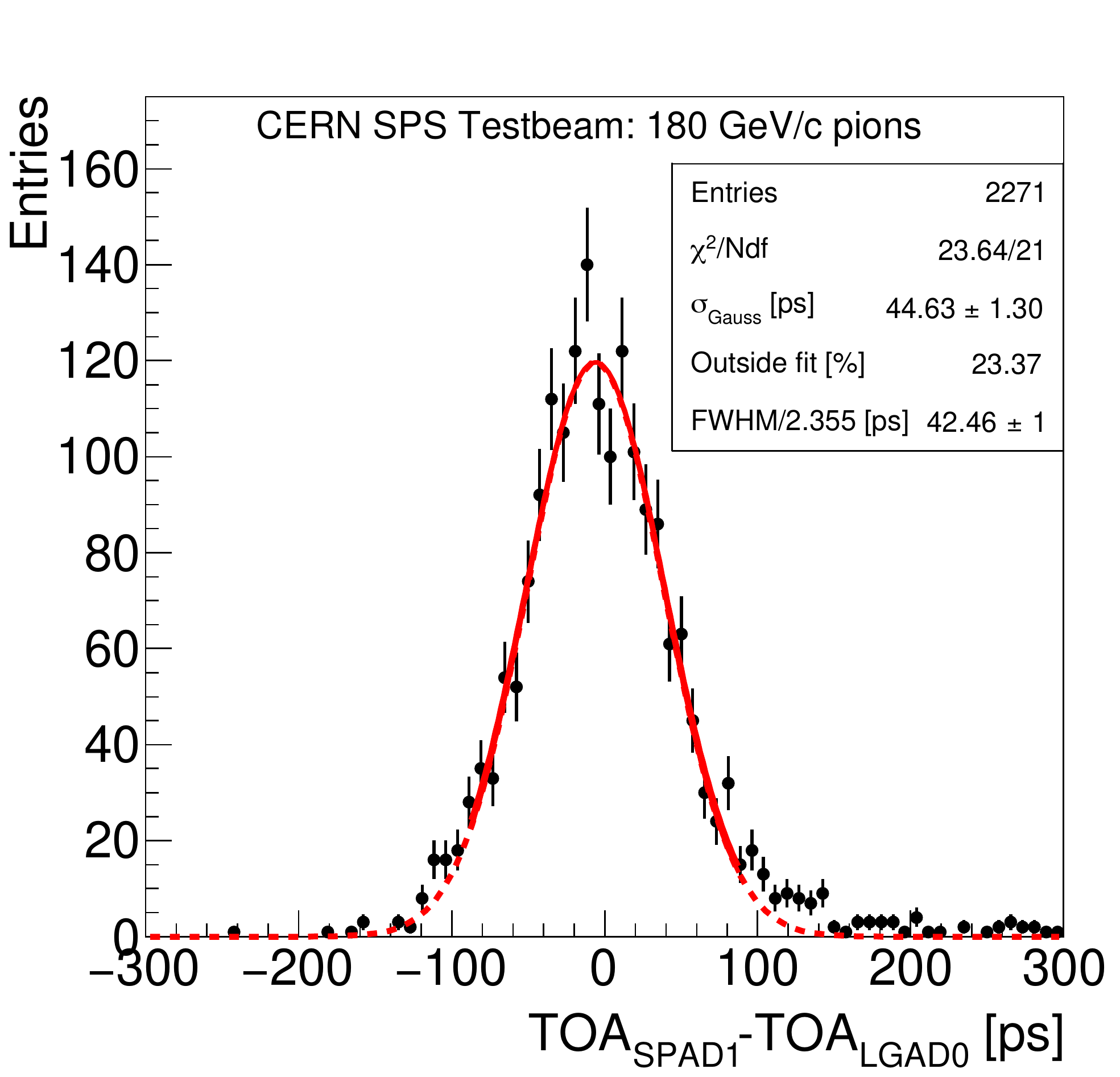}
\caption{\label{fig:TOF_spads_lgdads} $\dtoa$ difference between SPAD0 and LGAD0 (left), and between SPAD1 and LGAD0 (right) after time-walk correction for the LGAD0. The red lines show the results of a Gaussian fit to each distribution: the full line extends across the fit range considered, while the dashed line extrapolates the fitted function to the whole histogram.}
\end{figure}

An additional time jitter affects the $\dtoa$ measured between the two LGADs, connected to the first oscilloscope, and pixels OA1, OA2 and OA3, connected instead to the second oscilloscope. This extra jitter $\sigma_{{\it scopes}}$ stems from, for instance, the different bandwidth and sampling rate of the two oscilloscopes and the longer and lower-quality cables used during the testbeam measurements to connect the detectors to the second oscilloscope. Conversely, being read by the same oscilloscope as the LGADs, pixel OA0 does not suffer from this jitter. The corresponding  degradation in the measured time resolution for DUT pixels OA1, OA2 and OA3 was estimated using the two SPADs~\cite{charbon}. The distributions of the $\dtoa$ between LGAD0 and the SPAD0, which was read by the first oscilloscope, and between LGAD0 and the SPAD1, which was read by the second oscilloscope, were fit with a Gaussian function, as illustrated in Figure~\ref{fig:TOF_spads_lgdads}. 
The jitter produced by the two oscilloscopes was then estimated by subtracting in quadrature the two fitted $\sigma$ parameters, under the assumption of an identical time response between the two SPADs. 
This procedure provided  an estimated additional contribution of $\sigma_{{\it scopes}}=(16.1\pm4.6)~\si{\pico\second}$ from the second oscilloscope to the measured time resolution of DUT pixels OA1, OA2 and OA3.


\subsection{Results}

To measure the time resolution, a Gaussian fit  was performed simultaneously to all the $\dtoa$ distributions obtained with the four DUT pixels and the two LGADs, after the time-walk corrections, minimising a single $\chi^2$ function to extract directly the four DUT pixels and the LGAD resolutions $\sigdut{_{\it 0,1,2,3}}$, $\siglgzero$ and $\siglgone$. The $\sigma$ parameters of the Gaussian functions describing the $\dtoa$ distributions in the fit were computed using equations~\ref{eq:system}. For the DUT, a separate time resolution parameter was assigned to each pixel. 

In a second method, used as a cross check, the time resolution of each of the four DUT pixels was estimated separately. For each DUT pixel, a Gaussian fit to the three $\dtoa$ distributions was deployed to extract the $\sigdtoa{_{({\it DUT,LGAD0})}}$, $\sigdtoa{_{({\it DUT,LGAD01})}}$, and $\sigdtoa{_{({\it LGAD0,LGAD1})}}$ terms in the left hand side of Equations~\ref{eq:system}. The DUT and LGADs time resolutions were then computed afterward, by solving the system. The results obtained with the two methods were found to be compatible within statistics.

In all cases, the Gaussian fits were restricted to the bulk of each $\dtoa$ distribution by considering only bins with a number of entries larger than 25\% of the entries in the most populated bin. For all working points analyzed, the fraction of events exceeding the Gaussian fit integral  was always below 5\%, indicating that non-Gaussian contributions to the timing capabilities of the PicoAD prototype have overall a small impact, and that the resolutions quoted in the following refer to at least 95\% of the signals acquired.
\begin{figure}[!htb]
\centering %
\includegraphics[width=.32\textwidth,trim=0 0 0 0]{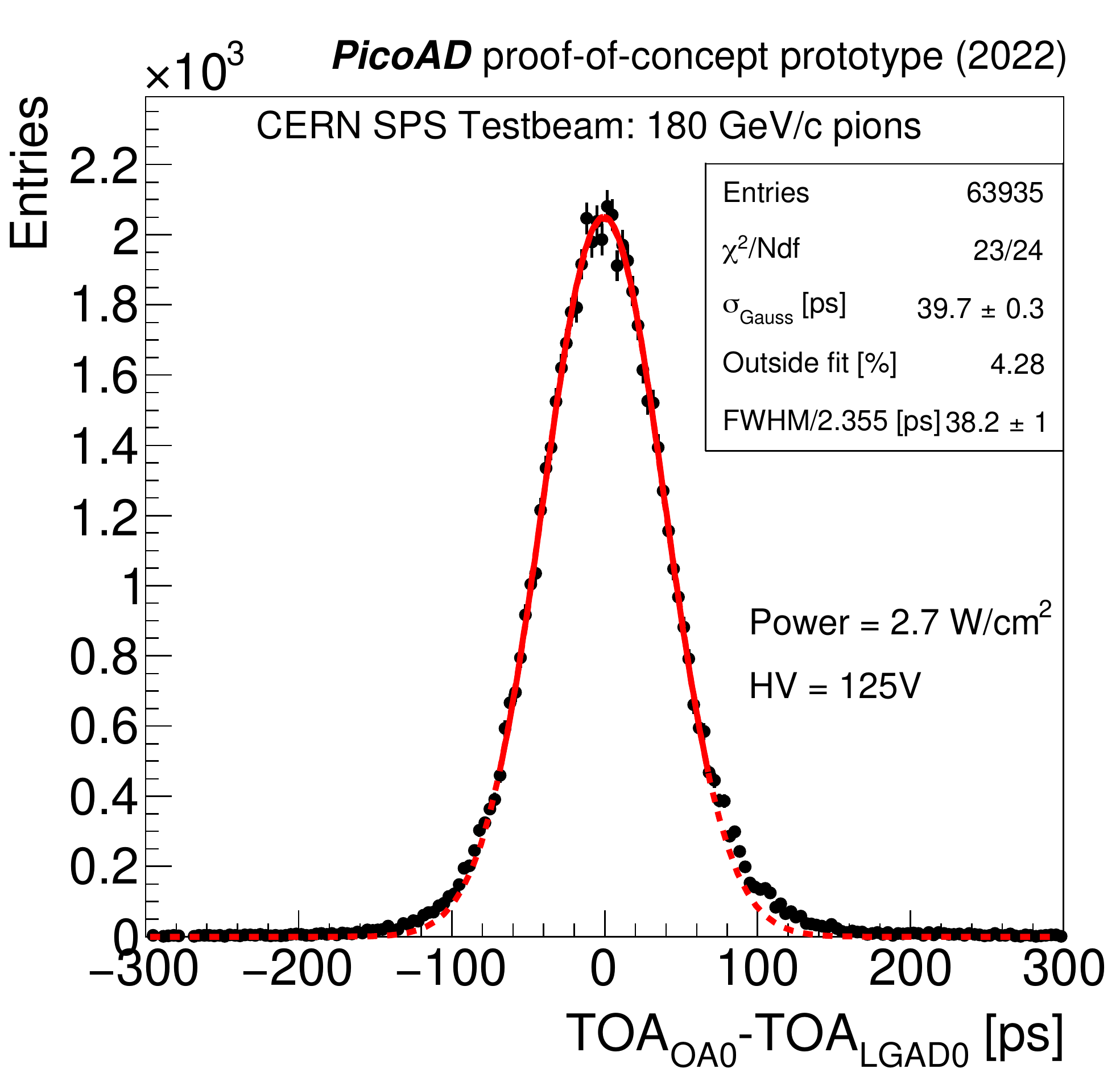}
\includegraphics[width=.32\textwidth,trim=0 0 0 0]{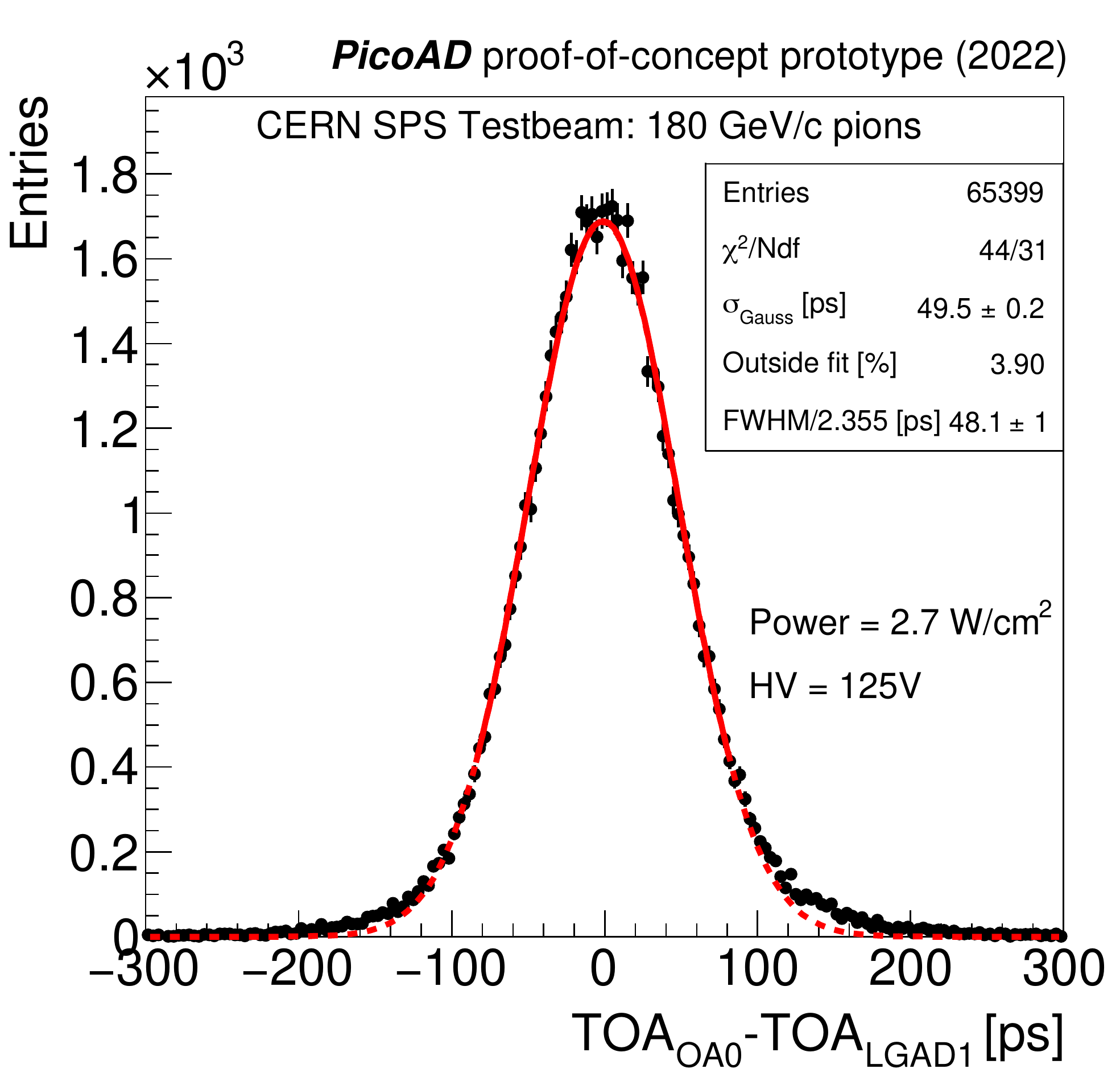}
\includegraphics[width=.32\textwidth,trim=0 0 0 0]{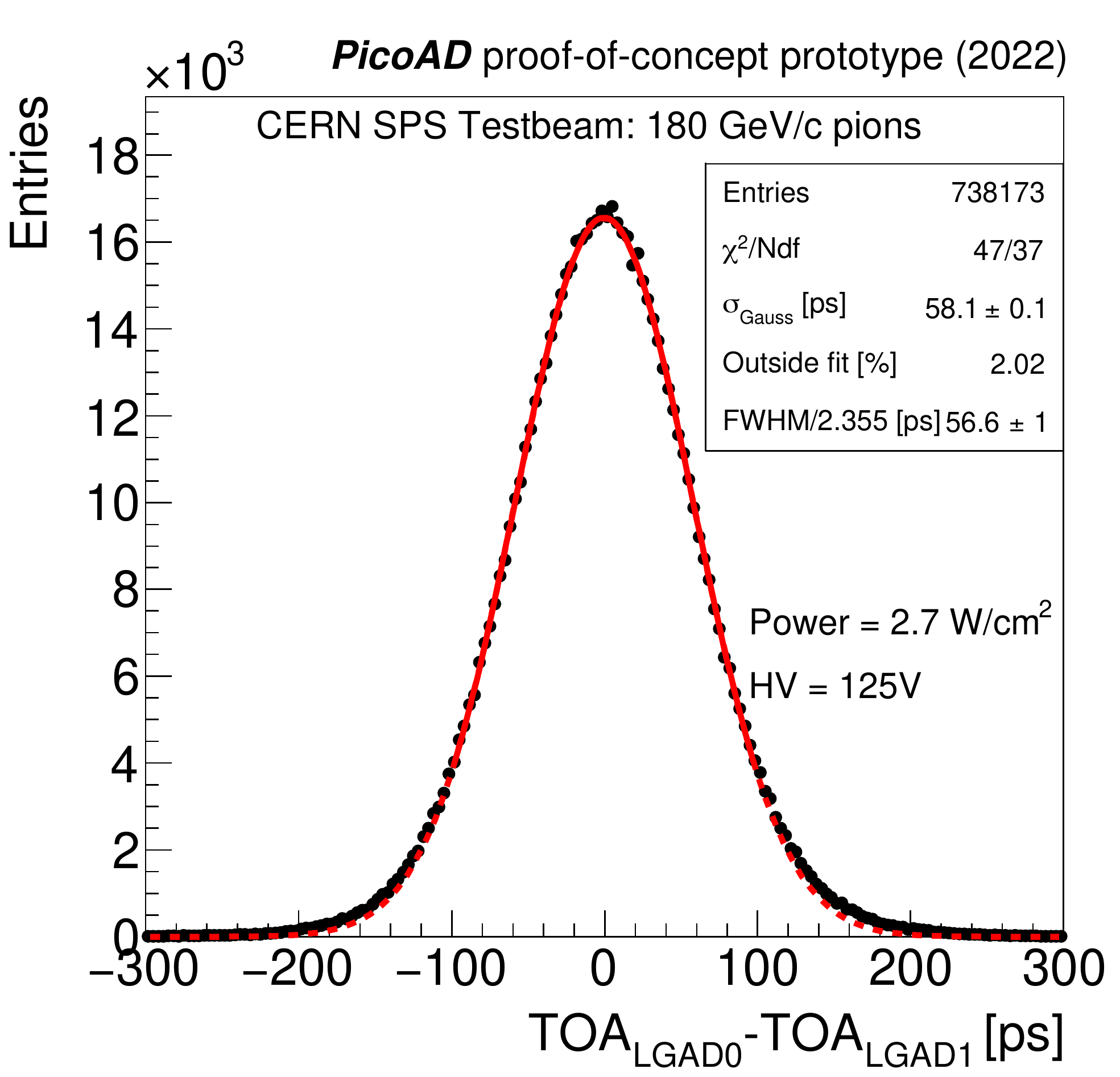}
\caption{\label{fig:TOF_fits} $\dtoa$ difference between pixel OA0 of the DUT and LGAD0 (left),  pixel OA0 of DUT and LGAD1 (center), and between LGAD0 and LGAD1 (right) after time-walk correction for the working point with $ \ipreamp = \SI{150}{\micro\ampere} $ (corresponding to a power density of 2.7 W/cm$^2$) and $ HV = \SI{125}{\volt} $. The red lines show the results of a Gaussian fit to each distribution: the full line extends across the fit range considered, while the dashed line extrapolates the fitted function to the whole histogram.}
\end{figure}

As an example, the three time-walk corrected $\dtoa$ distributions between DUT pixel OA0, LGAD0 and LGAD1 are shown in Figure~\ref{fig:TOF_fits} in the case  the PicoAD prototype was operated at a power density of 2.7 W/cm$^2$ 
and $ HV = \SI{125}{\volt} $.
The distribution of the $\dtoa$ between LGAD1 and the DUT is clearly larger than that involving LGAD0 and the DUT, which indicates that LGAD0 has a better time resolution than LGAD1. 
In addition, the $\dtoa$ distribution between the two LGADs is much larger than the other two $\dtoa$ distributions that contain the DUT, which shows that the DUT has a much better resolution that the two LGAD detectors used for this measurement.
The results of the Gaussian fit consist indeed of resolutions of
$(17.3\pm0.4)$ ps for  pixel OA0 of the DUT,
$(33.5\pm0.1)$ ps for LGAD0 and
$(41.7\pm0.1)$ ps for LGAD1.

The results for the four analog PicoAD pixels at this working point are summarized in Table~\ref{tab:resolution} for different 
conditions:
\begin{itemize}
\item The second column of the table shows the results in the case in which 
the DUT was not time-walk corrected at all (while the LGADs were corrected).
Even in this extreme case, the four DUT pixels  are able to provide  a time resolution at the level of 50 ps, which is very promising for application with e.g. very large number of readout channels.
\item In the case in which the time walk was corrected  using the signal amplitudes acquired with the oscilloscopes (third column of the table), the measured time resolutions vary between 24 ps in the case of pixel OA0 that was read by  the same oscilloscope of the LGADs, and approximately 30 ps for the other three pixels that were instead read out by the second oscilloscope and therefore are subject to the $\sigma_{{\it scopes}}$ jitter discussed in 
Section~\ref{oscjitter}.
\item Finally, in the case of the CFD-like time-walk correction explained in Section~\ref{sec:twc}, a time resolution of
$(17.3\pm0.4)$ ps was measured for pixel OA0.
The other three pixels,  read instead by the second oscilloscope and therefore suffering from the additional time jitter  $\sigma_{{\it scopes}}=(16.1\pm4.6)~\si{\pico\second}$, have a time resolution ranging from 24.4 to 25.3 ps.
Subtraction in quadrature of $\sigma_{{\it scopes}}$  provides the time-resolution values for pixels OA1, OA2 and OA3 reported in the last column of the table; these values are compatible with the result of pixel OA0, although within the relatively-large uncertainties coming from  the limited statistics that could be collected with the SPADs because of their small surface.
\end{itemize}


\begin{table}[!htb]
\centering
\renewcommand{\arraystretch}{1.4}
\begin{tabular}{|c|ccc|c|c|}
\cline{1-6}
\multicolumn{6}{|c|}{Time resolution $\sigma_{t} [\si{\pico\second}]$ for the DUT operated at $P_{\mathrm {\it density}} =$ 2.7 W/cm$^2$ and $ HV = \SI{125}{\volt} $} \\ 
\cline{1-6}

 \multirow{2}{*}{DUT pixel} & \multicolumn{3}{c|}{Time-walk correction: } &   \multirow{2}{*}{$\sigma_{{\it scopes}} $} &  \multirow{2}{*}{$\sqrt{\sigma_{{\it t,CFD-like}}^2-\sigma_{{\it scopes}}^2} $}\\ 
 \cline{2-4}
& no correction & amplitude-based & CFD-like &  & \\
\cline{1-6}
 \multicolumn{1}{|c|}{OA0}  &  49.3$\pm$1.5 & 24.1$\pm$0.4 &  17.3{\boldmath{$\pm$} }0.4 & \multicolumn{1}{c|}{-} & {\bf 17.3$\pm$0.4}\\ 
 \cline{1-6}
 \multicolumn{1}{|c|}{OA1}  & 49.7$\pm$1.5 & 30.5$\pm$0.3 &  25.3$\pm$0.3  & \multicolumn{1}{c|}{16.1$\pm$4.6} & {\bf 19.5$\mathbf{ \pm}$3.8}\\ 
 \multicolumn{1}{|c|}{OA2}  & 47.1$\pm$1.7 & 27.6$\pm$0.3 &  24.4$\pm$0.3  & \multicolumn{1}{c|}{16.1$\pm$4.6} & {\bf 18.3$\pm$4.1}\\ 
 \multicolumn{1}{|c|}{OA3}  & 47.5$\pm$1.6 & 29.4$\pm$0.3 &  24.8$\pm$0.3  & \multicolumn{1}{c|}{16.1$\pm$4.6} & {\bf 18.9$\pm$3.9}\\ 
 \cline{1-6}
\end{tabular}
\caption{Time resolution of the four analog pixels of the DUT for the working point at power density of 2.7 W/cm$^2$ and HV = 125  V for three cases: no time-walk correction; time-walk correction based on the signal amplitude; CFD-like time-walk correction. 
The last two columns show the  time jitter $\sigma_{{\it scopes}}$ between the two oscilloscopes measured using the SPADs, and the  time resolution of pixels OA1, OA2 and OA3 obtained  after $\sigma_{{\it scopes}}$ is subtracted in quadrature from the  time resolution obtained with the CFD-like time-walk correction, respectively. Only the statistical uncertainties are reported. }
\label{tab:resolution}
\end{table}

\begin{figure}[!htb]
\centering %
\includegraphics[width=.49\textwidth,trim=0 0 0 0]{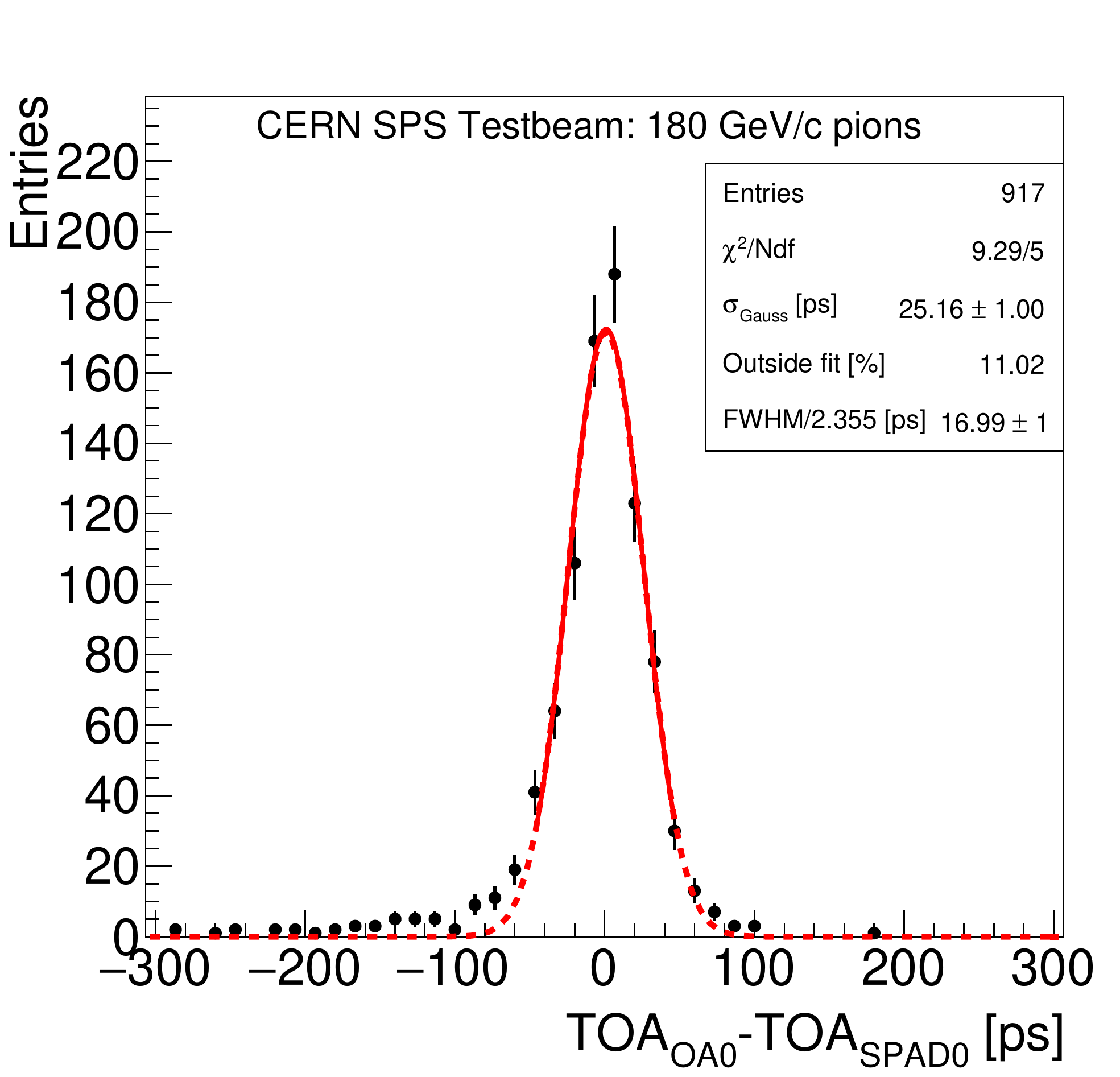}
\includegraphics[width=.49\textwidth,trim=0 0 0 0]{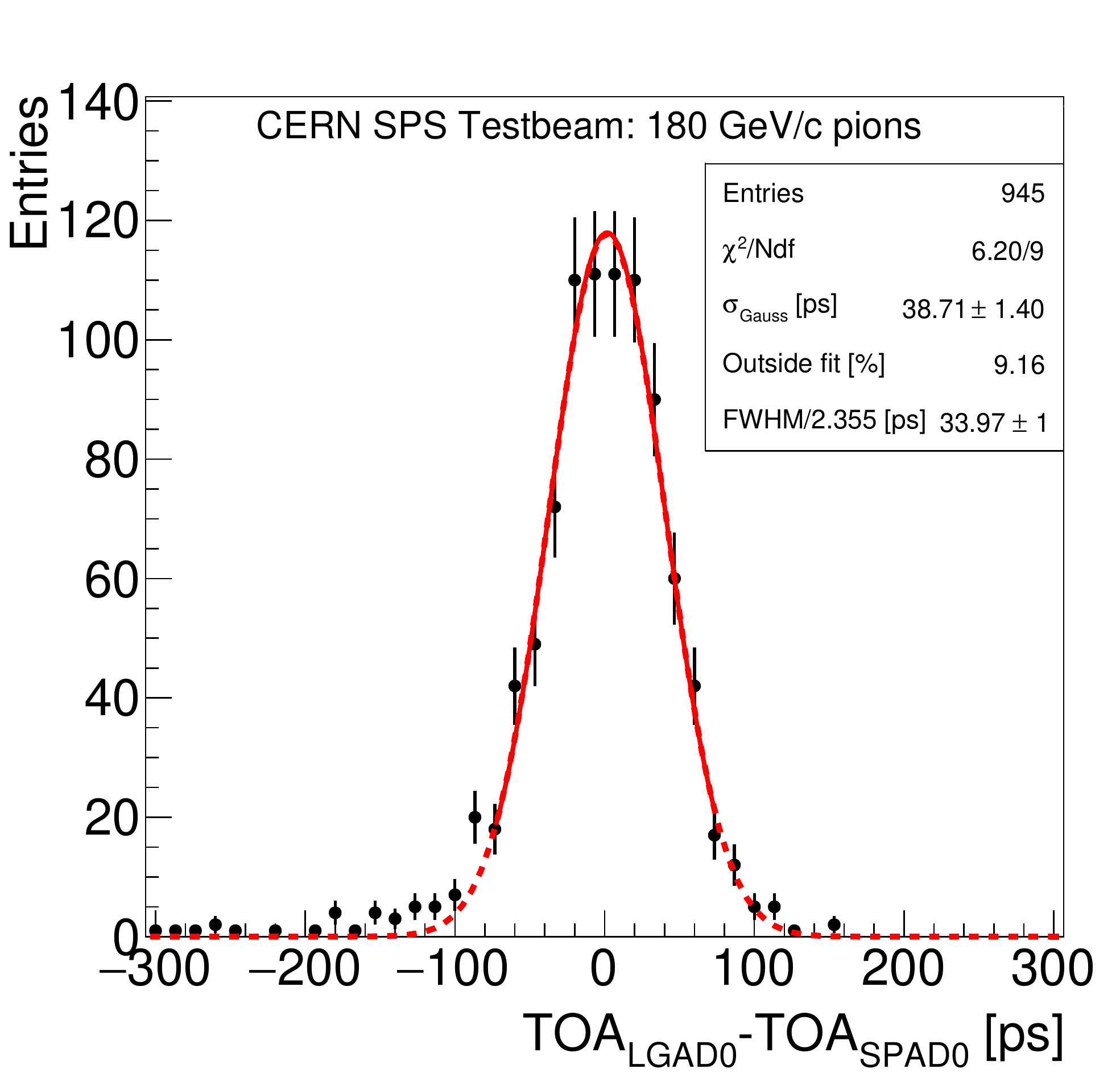}
\caption{\label{fig:TOF_fits_SPAD0} $\dtoa$ difference between pixels OA0 of the DUT and SPAD0 (left), and between LGAD0 and SPAD0 (right), after time-walk correction for the working point with $ \ipreamp = \SI{150}{\micro\ampere} $ (corresponding to a power density of 2.7 W/cm$^2$) and $ HV = \SI{125}{\volt} $. The red lines show the results of a Gaussian fit to each distribution: the full line extends across the fit range considered, while the dashed line extrapolates the fitted function to the whole histogram.
}
\end{figure}

These  time resolutions  are obtained constraining the resolution of the DUT using those of the LGADs that for some working points result to be worse than the DUT resolution.
To verify that this method does not produce a bias, we have repeated the data analysis using the pixel OA0 of the DUT together with the LGAD0 and substituting the LGAD1, which provides worst time resolution  than the LGAD0, with the SPAD0 that is read out by the same scope and therefore is not affected by $\sigma_{{\it scopes}}$. Given the much smaller  active area of the SPAD with respect to the LGAD active area, the data sample with the SPAD0 in the coincidence with the DUT and the LGAD0 is much smaller and thus the time resolution measurement suffers from a  larger statistical uncertainty. Figure~\ref{fig:TOF_fits_SPAD0} shows the distributions of the $\dtoa$ between the DUT pixel OA0 and SPAD0, and between the LGAD0 and SPAD0. 
The time resolution for the DUT pixel OA0 measured with this method amounts to $(17.1\pm3.5)$ ps and is thus compatible within statistical uncertainties with the measurement obtained using the two LGADs. 

\begin{figure}[!htb]
\centering %
\includegraphics[width=.49\textwidth,trim=0 0 0 0, clip]{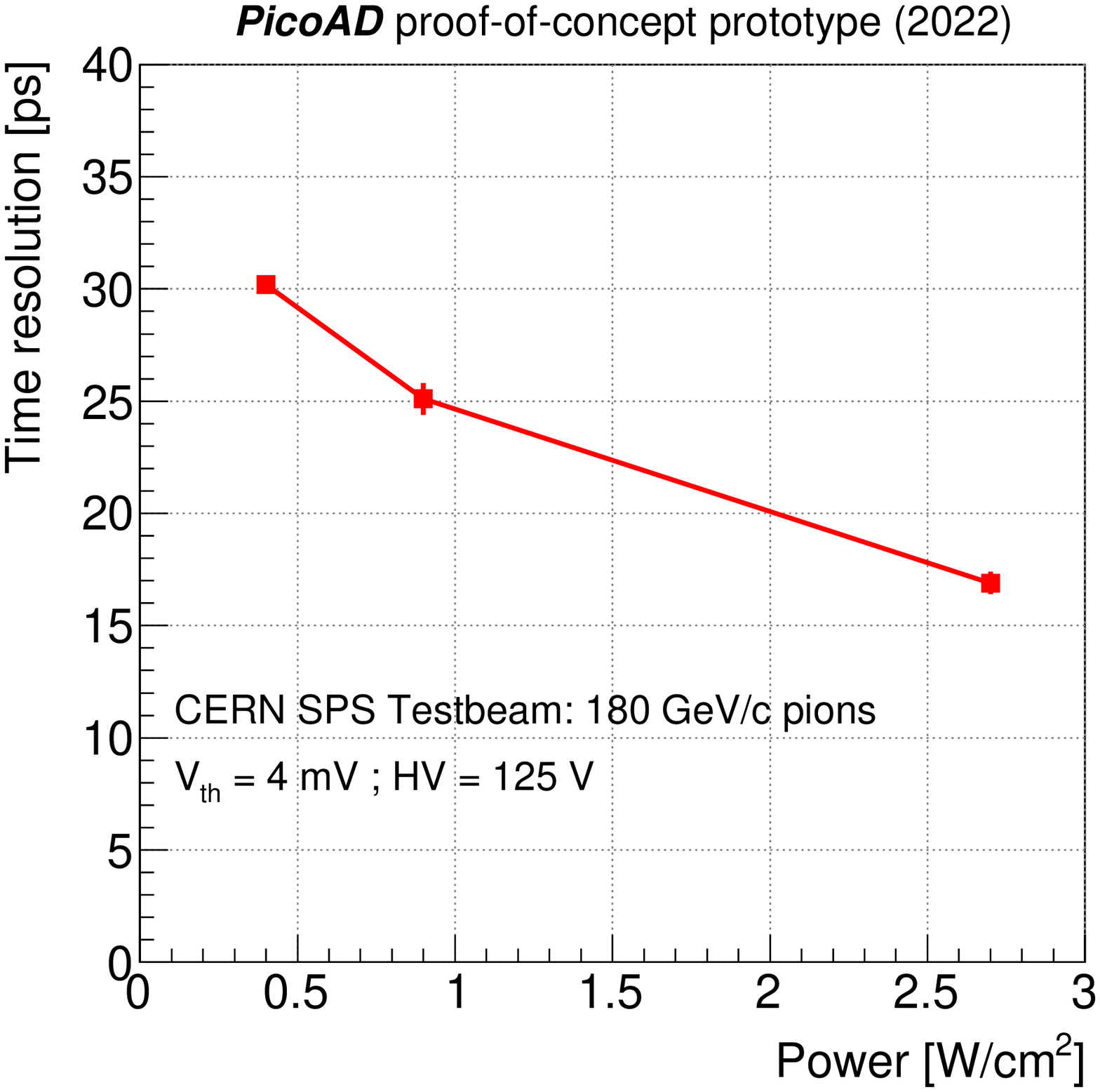}
~~\includegraphics[width=.49\textwidth,trim=0 0 0 0, clip]{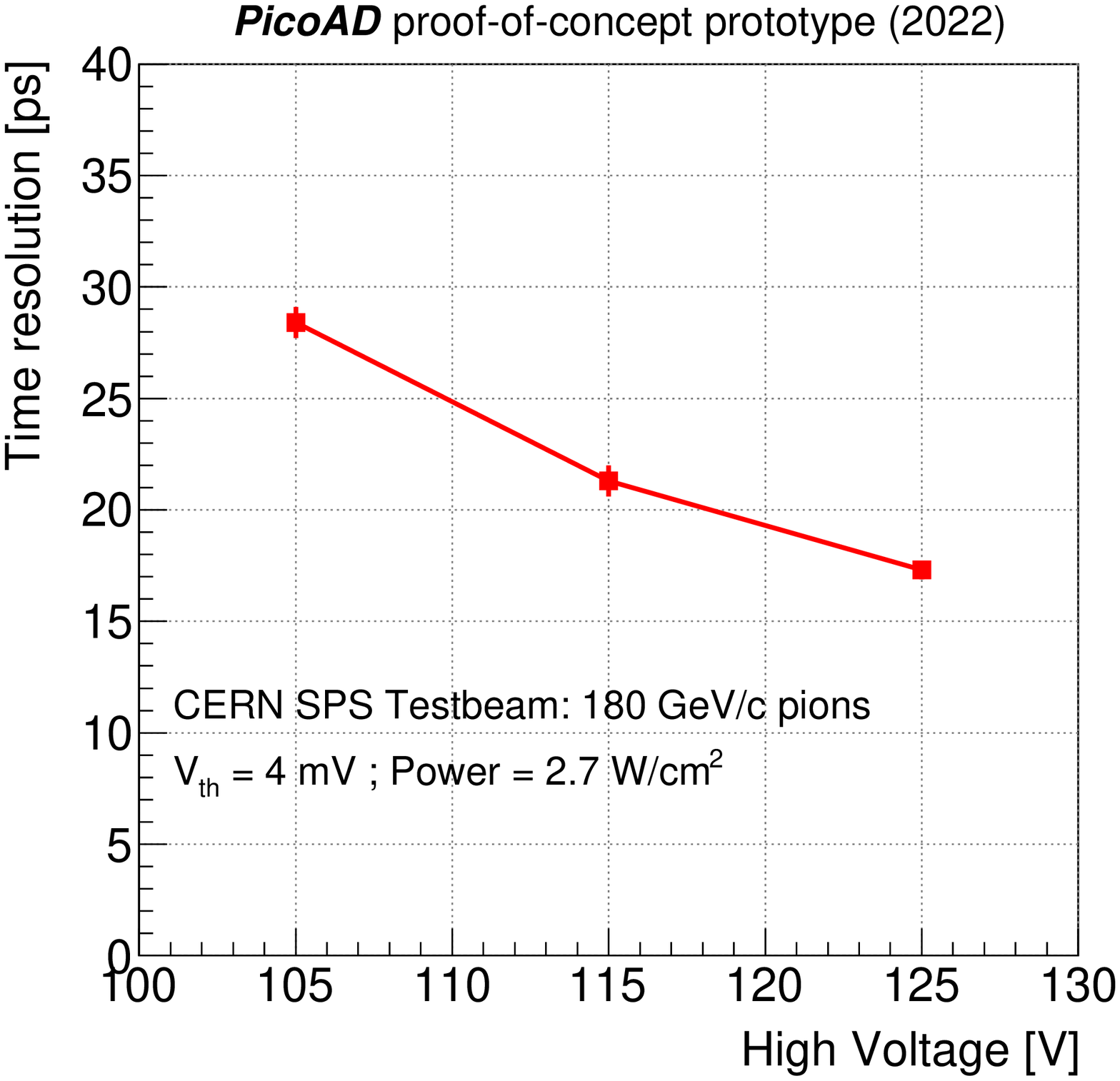}
\caption{\label{fig:TOFIPREAMP_HV} Time resolution for the  pixel OA0 of the DUT as a function of the power density at sensor bias voltage $HV = \SI{125}{\mV}$ (left panel), and as a function of sensor bias voltage at a power density of 2.7 W/cm$^2$ (right panel).
}
\end{figure}

Figure~\ref{fig:TOFIPREAMP_HV} left shows the time resolution for the DUT pixel OA0 measured at $HV=125$~V as a function of the amplifier power consumption per unit surface  $P_{\mathrm {\it density}}$. 
The trend indicates a progressive small deterioration of the timing performance of $(17.3\pm0.4)$ ps measured at 
 $P_{\mathrm {\it density}} =$ 2.7 W/cm$^2$
when the power consumption is reduced, although even at 0.4 W/cm$^2$ the time resolution remarkably remains  $\SI{30}{\pico\second}$. 

Figure~\ref{fig:TOFIPREAMP_HV} right  reports the results of the scan in sensor bias voltage  at $P_{\mathrm {\it density}} =$ 2.7 W/cm$^2$. The PicoAD proof-of-concept prototype provides a resolution of $(28.4 \pm 0.7)$ ps  at 105 V, which demonstrates 
that the detector can be operated at time resolutions better than 30 ps with a 20 V plateau in sensor bias voltage.

\begin{table}[!htb]
\centering
\renewcommand{\arraystretch}{1.3}
\begin{tabular}{|c|c|c|c|}
\cline{1-4}
\cline{1-4}
$\ipreamp$ [µA] & $P_{\it density}$ [W/cm$^2$] & Efficiency [\%] & Time Resolution [ps] \\
\cline{1-4}
 20 & 0.4  &  $ 99.90_{-0.05}^{+0.04} $ & $30.2 \pm 0.3$  \\
 50 & 0.9  &  $ 99.95_{-0.09}^{+0.04} $ & $25.1 \pm 0.7$  \\
150 & 2.7  &  $99.88_{-0.05}^{+0.04} $ & $17.3\pm 0.4$  \\
\cline{1-4}
\end{tabular}
\caption{Power consumption per unit surface, average efficiency calculated in the two triangles of Figure~\ref{fig:effmap} and pixel OA0 time resolution  measured at  $HV$ = 125 V  for the three pre-amplifier current values reported in the first column. }
\label{tabsumm} 
\end{table}

Table~\ref{tabsumm}
summarizes the efficiency and time resolutions 
obtained at the highest sensor bias $HV$ = 125 V for the three values of power consumption per unit surface considered during the testbeam.

\begin{figure}[!htb]
\centering %
\includegraphics[width=.49\textwidth,trim=0 0 0 0, clip]{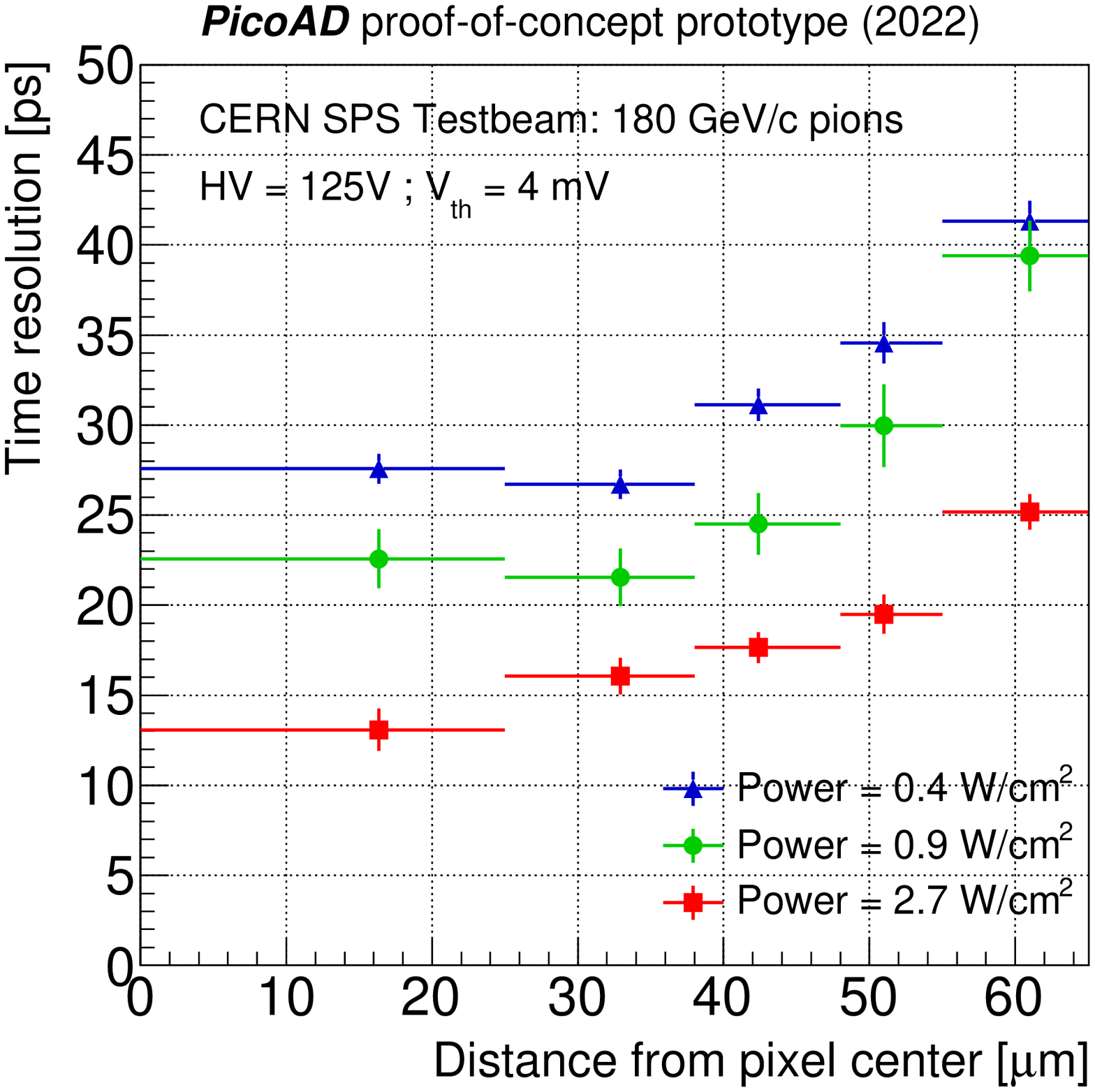}
~~\includegraphics[width=.48\textwidth,trim=0 0 0 0, clip]{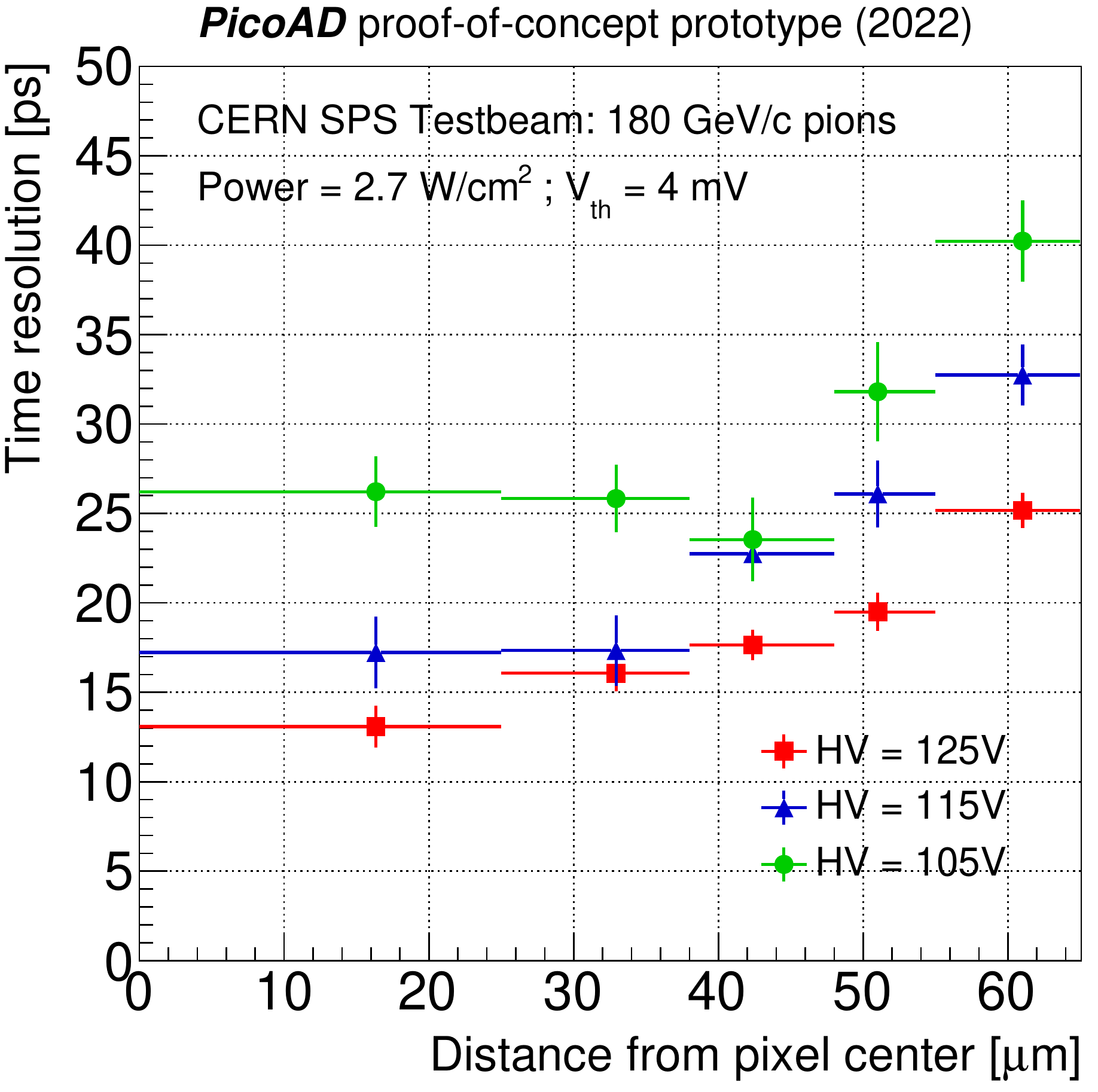}
\caption{\label{fig:TOF_RADIUS} 
Time resolution for the DUT pixel OA0 as a function of the distance from the pixel center. In the left panel, the  resolution is measured at $HV =$ 125 V for the three values of the power density considered, while in the right panel it is measured at a power density of 2.7 W/cm$^2$ for the three values of $HV$.
In each  bin, the data points are positioned at the mean value of the distance from the center of the pixel.}
\end{figure}

The time resolution was also studied as a function of the distance from the pixel center.
The results are shown in Figure~\ref{fig:TOF_RADIUS} left for $HV =$ 125 V for the three values of $P_{\it density}$ considered, while Figure~\ref{fig:TOF_RADIUS} right shows the resolutions obtained for $HV =$ 105, 115 and 125 V for $P_{\it density}=$ 2.7 W/cm$^2$. For the working point $HV$ = 125 V and $P_{\it density}=$ 2.7 W/cm$^2$ the data indicate  a time resolution of 13 ps at the center of the pixel and 25 ps for signals associated to tracks that point to the edge of the pixel. Similar relative trends between the center and the edge of the pixel are observed for the other working points studied.

A dedicated TCAD simulation shows that the lower time resolution at the pixel edge is explained partly by the lower signal amplitudes associated to charge-sharing between adjacent pixels induced by pions crossing the inter-pixel regions of the DUT, and partly by  the difference  in charge gain 
that   in this proof-of-concept prototype is lower toward the edge of the pixel with respect to the pixel center~\cite{picoad_gain}. 

%% file: Conclusions.tex

\section{Conclusions}
\label{sec:conclusions}

A  proof-of-concept  prototype of the PicoAD monolithic silicon detector was produced in 130 nm SiGe BiCMOS technology  with hexagonal pixels with $\SI{100}{\um}$ pitch. 
The prototype,  characterized by a fully-depleted double-junction (NP)$_{\it pixel}$(NP)$_{\it gain}$ structure, was tested at the CERN SPS facility with a beam of \SI{180}{\giga\electronvolt}/c momentum pions. 
Analog signals with amplitudes above a threshold of $\SI{4}{\mV}$, which corresponds to 7.5 times the voltage noise,
were retained for data analysis. 

At a sensor bias voltage $HV = 125$ V, the detection efficiency is measured to be compatible with 99.9\% when the power consumption per unit surface is varied between 0.4 and 2.7 W/cm$^2$.  

In this first prototype the time resolution depends significantly on the distance from the center of the pixel:  at a power consumption of 2.7 W/cm$^2$ and $HV$ = 125 V the time resolution varies between 13 ps at the center of the pixel and 25 ps at the edge of the pixel, while
in the full pixel area, including the inter-pixel regions, it is measured to be $(17.3\pm0.4)$ ps. 
At a reduced power consumption of 0.9 and 0.4 W/cm$^2$, the overall-pixel time resolution still remains $(25.1\pm0.7)$ and $(30.2\pm0.3)$ ps, respectively.

These results show that the  PicoAD concept works and that the addition of a continuous gain layer in a second PN junction far from the pixel matrix provides a fully efficient monolithic sensor at an affordable power consumption, that is able to  improve significantly the already remarkable performance obtained with a SiGe BiCMOS frontend in a monolithic sensor without a gain layer~\cite{Iacobucci:2021ukp}. 

Future PicoAD prototypes foreseen in the framework of the MONOLITH H2020 ERC Advanced project  will feature smaller pixel pitch,  reduction of the inter-pixel gap, further design improvements on the electronics  and overall reduction of the sensor capacitance. All these factors might further  increase the timing performance measured with this proof-of-concept  prototype, while maintaining operating capabilities with very high signal-to-noise ratios and  full detection efficiency at a contained power consumption.